\documentclass[twocolumn, twocolappendix]{aastex631}
\usepackage{amsmath}

\usepackage{gensymb}
\usepackage{xcolor}
\usepackage{comment}
\usepackage{makecell}
\usepackage{rotating}
\usepackage{float}
\newsavebox{\warptablebox}
\makeatletter
\renewcommand{\tablecomments}[1]{\par\noindent #1}
\makeatother

\makeatletter
\renewcommand{\@makefntext}[1]{%
  \hsize=\columnwidth
  \noindent
  \makebox[0pt][r]{\@makefnmark\hspace{0.5em}}#1%
}
\def\footnoterule{%
  \kern-3\p@
  \hrule width 0.7\hsize height 0.5pt
  \kern2.6\p@
}
\makeatother

\definecolor{MyDarkBlue}{rgb}{0,0.08,0.5}
\definecolor{MyDarkRed}{rgb}{0.7,0.02,0.02}
\definecolor{MyDarkmagenta}{rgb}{0.0,0.7,0.0}

\usepackage{comment}
\usepackage{tabularx}
\usepackage{booktabs}

\newcommand{\rev}[1]{#1}

\shortauthors{Ishibashi et al.}

\begin{document}

%% you desire. In v6.31 you can include a footnote in the title.

%% A significant change from earlier AASTEX versions is in the structure for 
%% calling author and affiliations. The change was necessary to implement 
%% auto-indexing of affiliations which prior was a manual process that could 
%% easily be tedious in large author manuscripts.
%%
%% The \author command is the same as before except it now takes an optional
%% argument which is the 16 digit ORCID. The syntax is:
%% \author[xxxx-xxxx-xxxx-xxxx]{Author Name}
%%
%==============================================================================
% Documents
%==============================================================================

%------------------------------------------------------------------------------
% Title
%------------------------------------------------------------------------------
\title{A Warped Dust Ring in the CQ Tau Disk}

%------------------------------------------------------------------------------
% Authors
%------------------------------------------------------------------------------
%% Affiliation
% New commands for affiliation
\newcommand{\asiaa}{Academia Sinica Institute of Astronomy and Astrophysics, 11F of ASMA Building, No.1, Sec. 4, Roosevelt Rd., Taipei 106216, Taiwan}
\newcommand{\kagoshima}{Department of Physics and Astronomy, Graduate School of Science and Engineering, Kagoshima University, 1-21-35 Korimoto, Kagoshima, Kagoshima 890-0065, Japan}
\newcommand{\kanazawa}{Advanced Research Center for Space Science and Technology, Kanazawa University, Kakuma, Kanazawa, Ishikawa, 920-1192, Japan}
\newcommand{\kogakuin}{Division of Liberal Arts, Kogakuin University, 1-24-2 Nishi-Shinjuku, Shinjuku, Tokyo 163-8677, Japan}
\newcommand{\kyushu}{Department of Earth and Planetary Sciences, Faculty of Sciences, Kyushu University, Nishi-ku, Fukuoka 819-0395, Japan}
\newcommand{\naoj}{National Astronomical Observatory of Japan, National Institutes of Natural Sciences, 2-21-1 Osawa, Mitaka, Tokyo 181-8588, Japan}
\newcommand{\tokyoscience}{Department of Earth and Planetary Sciences, Institute of Science Tokyo, 2-12-1 Ookayama, Meguro, Tokyo 152-8551, Japan}

\correspondingauthor{Shu Ishibashi}
\email{k8174036@kadai.jp}
\author[0009-0001-1889-9216]{Shu Ishibashi}
\affiliation{\kagoshima}

\author[0000-0002-8185-9882]{Masayuki Yamaguchi}
\affiliation{\kyushu}
\affiliation{\naoj}

\author{Takayuki Muto}
\affiliation{\kogakuin}

\author[0000-0001-9304-7884]{Naomi Hirano}
\affiliation{\asiaa}
\affiliation{\kanazawa}

\author[0000-0003-4039-8933]{Ryuta Orihara}
\affiliation{\tokyoscience}

\author[0000-0003-1549-6435]{Kazuya Saigo}
\affiliation{\asiaa}

\author[0000-0003-1412-893X]{Hsi-Wei Yen}
\affiliation{\asiaa}

\author[0000-0003-0845-128X]{Shigehisa Takakuwa}
\affiliation{\kagoshima}


%This example manuscript is intended to serve as a tutorial and template for authors to use when writing their own AAS Journal articles. The manuscript includes a history of \aastex\ and includes figure and table examples to illustrate these features. Information on features not explicitly mentioned in the article can be viewed in the manuscript comments or more extensive online documentation. Authors are welcome replace the text, tables, figures, and bibliography with their own and submit the resulting manuscript to the AAS Journals peer review system.  The first lesson in the tutorial is to remind authors that the AAS Journals, the Astrophysical Journal (ApJ), the Astrophysical Journal Letters (ApJL), the Astronomical Journal (AJ), and the Planetary Science Journal (PSJ) all have a 250 word limit for the  abstract\footnote{Abstracts for Research Notes of the American Astronomical Society (RNAAS) are limited to 150 words}.  If you exceed this length the Editorial office will ask you to shorten it.

\begin{abstract}
The diversity of planetary orbital architectures calls for observational constraints on \rev{radial variations in disk-plane orientation, or warps, yet such variations} in dust disks remain poorly characterized. CQ~Tau is a promising target because its misaligned inner disk and candidate gas warp suggest a radially extended warp. We present the ALMA 1.3 mm dust continuum image reconstructed from the real part of the visibilities, extracting the centro-symmetric disk structure. At an effective spatial resolution of $0.052'' \times 0.042''$ ($7.8 \times 6.3$ au), the image reveals a bright main ring at radii of $\sim20$--$70$ au and an unresolved compact emission component \rev{tracing the inner disk around the central star}. Ellipse fits centered on the stellar position to contours tracing the inner and outer edges of the ring reveal radial variations in position angles and inclinations. At the inner edge ($\sim21$--$26$ au), the position angle and inclination span $214^\circ$--$233^\circ$ and $23^\circ$--$39^\circ$, respectively, while they span $231^\circ$--$235^\circ$ and $33^\circ$--$36^\circ$ at the outer edge ($\sim58$--$67$ au). Whereas a coplanar ring toy model fails to reproduce such variations, a warped ring model with the local tilt decreasing from $\sim4^\circ$ at $r = 20$ au to $\sim1^\circ$ at 60 au reproduces them. The high-pass-filtered real-part image reveals a pair of opposed curved negative residuals. Similar features appear only in the warped ring model, further supporting the dust warp. These results establish CQ~Tau as an example of a warped dust disk and an important laboratory for investigating the connection between three-dimensional disk geometry and planet formation.
\end{abstract}



\section{Introduction}
\label{sec:intro}
Exoplanet observations have shown that planetary systems exhibit a wide variety of orbital inclinations and misalignments \citep{Winn2015,Albrecht2022}. This diversity suggests that protoplanetary disks, the birthplaces of planets, should be understood not only as single planar structures but also as systems with three-dimensional geometries \citep{Stolker2016,Kraus2020}. High-angular-resolution observations with the Atacama Large Millimeter/submillimeter Array (ALMA) have revealed that protoplanetary disks commonly host a variety of substructures, such as rings, gaps, spiral arms, and azimuthal asymmetries \citep[e.g.,][]{Andrews2018,Long2018,Yamaguchi2024,Shoshi2025b}. 
More recently, warped disks, in which the disk rotation axis varies with radius, have been reported in several protoplanetary disks \citep[e.g.,][]{Marino2015,Casassus2018,Bi2020,Sai2020,Orihara2025}.
\rev{\citet{Bohn2022} constrained orientations of compact inner disks in 20 transition disks using near-infrared (NIR) interferometric observations and found significant misalignments between the inner and outer disks in six systems. Such misaligned inner disks can partially block stellar irradiation and cast shadows on the outer disks, providing an observational signature of disk misalignment \citep[e.g.,][]{Marino2015}. Theoretically, disk warps and/or misalignments can be induced by gravitational perturbations from companions \citep{Juhasz2017}, stellar flybys \citep{Kimmig2026}, misaligned infall \citep{Kuffmeier2021}, and magnetic star--disk interactions \citep{Foucart2011}.}
Such variations in the rotation axis suggest that disk structure should be characterized not only in terms of localized substructures but also in terms of how the disk orientation changes with radius. However, most of these warps have been identified through gas kinematics or NIR observations. The geometry of the dust disk should also be important, because dust grains constitute the solid reservoir from which planetesimals and planetary cores form. Establishing an observational framework to quantify dust-disk warps as radial variations in disk orientation is therefore essential for connecting disk geometry with planet formation.

CQ~Tau, a relatively evolved protoplanetary disk system with an estimated age of $\sim10~\rm{Myr}$ \citep{Garufi2024,Wallack2024} at a distance of 149.4~pc \citep{GaiaCollaboration2023}, is a particularly promising target for investigating dust disk warps.
This transition disk exhibits a variety of structures, including a bright dust ring, crescents, and arcs in the dust continuum, \rev{NIR shadows on the dust ring,} as well as spiral arms seen in NIR and gas observations \citep{UbeiraGabellini2019,Wolfer2021,Hammond2022,Ren2023,Yamaguchi2024,Curone2025,Zagaria2025}. 
In addition, a misaligned inner disk \rev{(with a tilt of $\sim40^\circ$) has been reported by \cite{Bohn2022}}. \citet{Winter2025} showed that the non-axisymmetric CO line-of-sight velocity pattern of CQ~Tau can be reproduced by the velocity field expected for a warped disk. 
The misaligned inner disk and the candidate gas warp suggest that CQ~Tau may host a warp extending over a wide radial range. Gaia astrometry has revealed a proper motion anomaly that is difficult to explain with a single-star model, suggesting the presence of a companion \citep{Blakely2026,Vioque2026}.

We combine three archival ALMA Band~6 datasets of CQ~Tau and reconstruct a 1.3 mm dust continuum image using the sparse-modeling-based two-dimensional super-resolution imaging method PRIISM \citep{Nakazato2020,Nakazato2020_SPIE}.
\rev{PRIISM performs visibility fitting under the assumptions of image sparsity and smoothness, and reconstructs images consistent with the observed visibilities. In contrast to the standard CLEAN algorithm, PRIISM does not require deconvolution and convolution procedures, providing the potential to achieve higher image fidelity and super resolution.}
Using this image, we characterize the geometry of the dust disk as a function of radius and test whether this geometry can be explained by a warp.

This paper is organized as follows. In Section~\ref{sec:data_reduction_and_imaging}, we describe the ALMA Band~6 data, data reduction, and CLEAN and PRIISM imaging procedures. In Section~\ref{sec:results}, we present the 1.3 mm dust continuum morphology revealed by the CLEAN, PRIISM, and real-part PRIISM images. In Section~\ref{sec:analysis}, we derive the global disk orientation and its radial variations through ellipse fitting and compare the results with those of coplanar-ring and warped-ring toy models. In Section~\ref{sec:discussion}, we discuss the evidence for the dust warp and its possible physical origins. Section~\ref{sec:summary} summarizes the main results and interpretations.

\section{Data Reduction and Imaging}
\label{sec:data_reduction_and_imaging}

\subsection{Data Reduction}
\label{sec:data_reduction}

We combined and analyzed three ALMA archival datasets of the 1.3 mm dust continuum emission of CQ~Tau, which consist of Data 1 (2013.1.00498.S), Data 2 (2016.A.00026.S), and Data 3 (2017.1.01404.S). Data 3 consists of two execution blocks, hereafter referred to as Data 3-1 and Data 3-2. These data were observed in 2015 August, 2017 August, and 2017 November, respectively. The 1.3 mm dust continuum data from these observations have already been published \citep{UbeiraGabellini2019,Wolfer2021}, and the configuration, baseline, and other observational properties of the datasets are summarized in \cite{Wolfer2021}. Each dataset was calibrated with the standard ALMA calibration pipeline, with Common Astronomy Software Applications \citep[CASA;][]{McMullin2007} versions 4.3.1, 4.7.2, and 5.1.1-5 for Data 1, Data 2, and Data 3, respectively. The further calibration described below was performed with CASA version 6.1.0.

To combine Data 1, 2, 3-1, and 3-2, we first aligned their phase centers using the CASA tasks \texttt{fixvis} and \texttt{fixplanets}. In this step, we made PRIISM continuum images separately for Data 1, Data 2, and Data 3, following the procedure described in Section~\ref{sec:imaging_with_CLEAN_and_PRIISM}. Here, Data 3-1 and Data 3-2 were imaged together because their observing dates were separated by only three days. These images were used to check whether the compact inner disk emission was detected and could serve as the reference position for phase alignment. For Data 1, since the compact inner emission was not detected, likely due to limited angular resolution, we adopted the ICRS coordinates R.A. = $05^{\rm h}35^{\rm m}58.4700^{\rm s}$ and Dec. = $+24^\circ44'53.683''$, corresponding to the average of the two stellar positions expected for 2015 and 2016 from Gaia DR3 astrometry.
\rev{Since Data 1 was obtained in 2015 August, the average stellar position obtained from the 2015 and 2016 Gaia observations, which corresponds to the position in the middle of 2015, should provide a sufficiently accurate stellar position. Assuming the linear proper motion, the offset between this averaged position and the stellar position at the actual observing epoch is at most $\sim5$ mas, much smaller than the effective spatial resolution of the PRIISM image for Data 1, $\sim0.12''$.}
For Data 2 and Data 3, the compact inner disk emission was detected in the PRIISM images, and its peak position was adopted as the reference position. The adopted coordinates were R.A. = $05^{\rm h}35^{\rm m}58.4697^{\rm s}$ and Dec. = $+24^\circ44'53.615''$ for Data 2, and R.A. = $05^{\rm h}35^{\rm m}58.4710^{\rm s}$ and Dec. = $+24^\circ44'53.618''$ for Data 3. The peak-based positions are close to the stellar position expected for 2018 from Gaia DR3 astrometry, with offsets of $0.011''$ for Data 2 and $0.0067''$ for Data 3. The position adopted for Data 3 was applied to both Data 3-1 and Data 3-2. Because \texttt{fixplanets} accepts directions only in the J2000 frame, the adopted ICRS positions were converted to J2000 before applying \texttt{fixvis}. The J2000 coordinate of the Data 3 position was then assigned as the common phase-center coordinate for all datasets using \texttt{fixplanets}. Accordingly, we use this common phase center as a proxy for the stellar position throughout this study.

After aligning the phase centers, we adjusted their flux scales. Specifically, we first deprojected the $uv$ coordinates assuming a position angle (PA) of $235^\circ$ and an inclination $i$ of $35^\circ$, following the disk geometry adopted by \citet{UbeiraGabellini2019}, and compared the azimuthally averaged real visibility profiles within overlapping deprojected $uv$-ranges to derive the relative flux-scaling factors. The real- and imaginary-part profiles show that the imaginary part remains close to zero, and the null positions in the real part associated with the ring structure are consistent among the datasets. Here, Data 2 was used as the reference for the flux scaling because it provides the most extensive coverage at short $uv$-spacings. The flux discrepancies relative to Data 2 were $35\%$, $3\%$, and $8\%$ for Data 1, Data 3-1, and Data 3-2, respectively. The corresponding antenna-based gain factors of 1.164, 1.014, and 1.040 were input to the CASA task \texttt{gencal} to generate the amplitude calibration tables, which were then applied to each dataset using \texttt{applycal}. The larger flux deviation of Data 1 may reflect an absolute flux-calibration offset in the oldest observations at Cycle 2.

Following the phase-center alignment and flux rescaling, all self-calibration steps described below were performed using the automated self-calibration pipeline \rev{\texttt{AJISAI v0.1.0} \citep{Yamaguchi2026AJISAI}}\footnote{\texttt{AJISAI} is publicly available at \url{https://github.com/Y-Masayuki/AJISAI}.}. 
\rev{A detailed description of \texttt{AJISAI} is provided in Appendix B of \cite{Yamaguchi2026}.}
Self-calibration was first performed independently for each dataset. Two rounds of phase self-calibration, with a longer solution interval for the first round and a shorter one for the second (Data 1: \texttt{inf} and 121 s; Data 2: \texttt{inf} and 40 s; Data 3-1: \texttt{inf} and 40 s; Data 3-2: \texttt{inf} and 40 s), were applied to improve the signal-to-noise ratio (SNR). The SNR was evaluated as the ratio of the peak intensity to the rms noise measured in an emission-free circular annulus between $1.2''$ and $2.0''$ from the phase center. These self-calibration steps improved the SNRs of the individual datasets by factors of 2.5, 2.0, 2.0, and 1.8, respectively. 

Finally, the combined data were self-calibrated with one round of phase self-calibration with a solution interval of \texttt{inf}. This final self-calibration improved the SNR of the combined dataset by a factor of 1.2 relative to its value before the final self-calibration.
% --- data combine coords
% ICRS
% data1: '5:35:58.4700', '24:44:53.683'
% data2: '5:35:58.4697', '24:44:53.615'
% data3-1,3-2: '5:35:58.4710', '24:44:53.618'

%====================================Fig====================================
\begin{figure*}[t]
\centering
\includegraphics[width=1.0\textwidth]{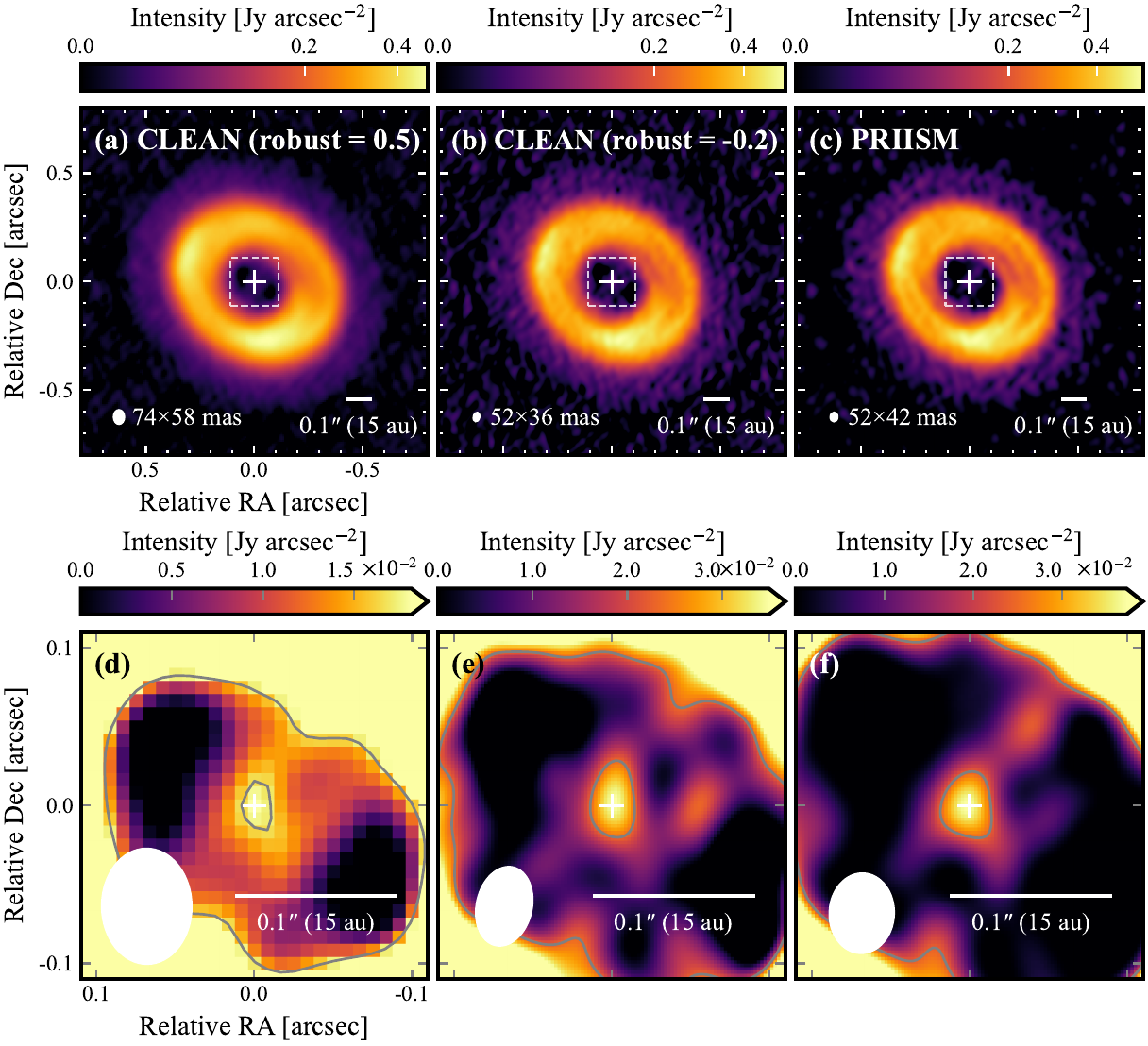}
\caption{%zoomin contour: clean(rb05)6sigma, clean(rbm025)3sigma, priism1IDT
CLEAN and PRIISM images of the ALMA 1.3 mm dust continuum emission in CQ~Tau. 
\rev{
Top panels show the entire picture of the disk with a color scale of a power-law exponent of 0.5, while bottom panels show zoomed-in views around the adopted phase center with a linear color scale. White dashed boxes in top panels indicate the zoomed-in regions. White crosses indicate the adopted phase center (i.e., the stellar position), and scale bars of $0.1''$ (15 au) are shown in all panels for reference. White ellipses indicate the synthesized beams for the CLEAN images and the effective spatial resolution for the PRIISM image, with their sizes labeled in panels (a)--(c).
(a) CLEAN image with the Briggs robust parameter of 0.5.
(b) CLEAN image with the robust parameter of -0.2. 
(c) PRIISM image. 
(d)--(f) Zoomed-in views corresponding to panels (a)--(c), respectively. Gray contours indicate the $6\sigma$, $3\sigma$, and $I_{\rm DT}$ levels, in panels (d), (e), and (f), respectively, where $\sigma$ is the rms noise measured for each CLEAN image.}
}
\label{fig:cleanpriism}
\end{figure*}
%====================================Fig====================================

\subsection{Imaging with CLEAN and PRIISM}
\label{sec:imaging_with_CLEAN_and_PRIISM}

After obtaining the combined visibility data, we reconstructed 1.3 mm dust continuum images using both CLEAN and PRIISM. We first produced a CLEAN image with Briggs weighting and \rev{a default} robust parameter of 0.5. The resulting CLEAN image has a synthesized beam of $0.074''\times0.058''$ with a PA of $0.6^\circ$. The rms noise was measured in an emission-free circular annulus between $1.2''$ and $2.0''$ from the phase center, yielding \rev{2.78}$~{\rm mJy~arcsec^{-2}}$ (\rev{13.6}$~\mu{\rm Jy~beam^{-1}}$). The peak intensity is \rev{480$~{\rm mJy~arcsec^{-2}}$}, corresponding to a peak SNR of \rev{173}. The flux density measured from pixels above the 5$\sigma$ threshold is \rev{137.9}$~{\rm mJy}$.

\rev{To enable a direct comparison with the PRIISM image, where the effective spatial resolution is $0.052''\times0.042''$ as described below, we also produced a CLEAN image with a robust parameter of $-0.2$. The resulting CLEAN image has a synthesized beam size of $0.052''\times0.036''$ with a PA of $167.4^\circ$. The rms noise is $7.71~{\rm mJy~arcsec^{-2}}$ ($16.5~\mu{\rm Jy~beam^{-1}}$), and the peak intensity is $515~{\rm mJy~arcsec^{-2}}$, corresponding to a peak SNR of $67$. The flux density measured from pixels above the $5\sigma$ threshold is $131.5~{\rm mJy}$.}

For the PRIISM reconstruction, we used the Python package \texttt{PRIISM} version 0.12.0 \citep{Nakazato2020,Nakazato2020_SPIE}, which enables super-resolution imaging with improved image fidelity \citep{Yamaguchi2020,Yamaguchi2021,Yamaguchi2024,Shoshi2025a,Shoshi2025b}.
The hyper-parameters in PRIISM, $\Lambda_l$ and $\Lambda_{\rm tsv}$, were selected using 10-fold cross-validation (CV) available in \texttt{PRIISM}. This procedure allows us to objectively identify the image with the highest fidelity. 
Following the CV procedure described in \citet{Yamaguchi2020}, the optimal PRIISM image was reconstructed using the hyper-parameter set $(\Lambda_{l}, ~\Lambda_{\rm tsv}) = (3\times10^{5},~4\times10^{13})$.
\footnote{
The hyper-parameter search was performed in a coarse-to-fine manner. 
We first searched a broad logarithmic grid over 
$\Lambda_{l}=10^{3}$--$10^{7}$ and 
$\Lambda_{\rm tsv}=10^{12}$--$10^{16}$. 
We then refined $\Lambda_{\rm tsv}$ over $10^{13}$--$10^{15}$ with 
$\Lambda_l$ fixed at $10^{5}$, and subsequently refined $\Lambda_l$ over 
$10^{4}$--$10^{6}$ with $\Lambda_{\rm tsv}$ fixed at $4\times10^{13}$.}
The hyper-parameters for the PRIISM images used for the phase-center alignment in Section~\ref{sec:data_reduction} were also determined using the same CV-based parameter-search procedure. The adopted hyper-parameter sets $(\Lambda_{l},\Lambda_{\rm tsv})$ were $(10^{4},10^{9})$, $(10^{5},10^{12})$, and $(2\times10^{5},7\times10^{13})$ for Data 1, Data 2, and Data 3, respectively.

For the PRIISM image, the detection threshold is defined in a different way from the rms noise commonly used for CLEAN images. This is because, unlike CLEAN images, PRIISM images are reconstructed through regularized optimization under a non-negativity constraint, and the resulting pixel values are spatially correlated. Therefore, following previous PRIISM studies \citep{Yamaguchi2021,Yamaguchi2024}, we define the detection threshold, $I_{\rm DT}$, as the maximum intensity in an emission-free region. This provides a conservative threshold for identifying emission above the reconstruction noise floor. The resulting $I_{\rm DT}$ for the optimal PRIISM image was $24.0~{\rm mJy~arcsec^{-2}}$. The flux density measured from pixels above $I_{\rm DT}$ within the source region is $136.2~{\rm mJy}$, comparable to \rev{the 137.9$~{\rm mJy}$ and 131.5$~{\rm mJy}$ measured from the CLEAN images with robust parameters of 0.5 and -0.2, respectively.}

Spatial resolution is not automatically defined in PRIISM images as it is in CLEAN, because PRIISM reconstructs images by minimizing a cost function without any beam-convolution process. We evaluated the effective spatial resolution of the PRIISM image, $\theta_{\rm eff}$, following the method of \citet{Yamaguchi2021}.
The method consists of three steps. 
First, we created four point-source-injected visibility datasets by adding an artificial point source to the observed visibility data at a distance of $1.2''$ from the phase center toward the north, south, east, and west. The flux density of the injected point source was $7.0~{\rm mJy}$, corresponding to approximately $5\%$ of the total flux of the PRIISM image.
Second, we performed PRIISM imaging for each point-source-injected dataset using the optimal hyper-parameters, $(\Lambda_{l}, \Lambda_{\rm tsv}) = (3\times10^{5},~4\times10^{13})$. 
Third, we fitted the injected point source in each resulting PRIISM image with a two-dimensional Gaussian function and defined $\theta_{\rm eff}$ from the mean FWHM of the fitted Gaussians. The four injection tests yielded consistent FWHM values, confirming that the estimated effective resolution is insensitive to the adopted injection positions. The resulting value was $\theta_{\rm eff} = 0.052'' \times 0.042''$ (or $7.8 \times 6.3$ au) with a PA of $177.6^{\circ}$. 
% The PRIISM image achieves an improvement in effective resolution by a factor of 1.4 as compared to that of the CLEAN image. The peak intensity of the PRIISM image was $531~{\rm mJy~arcsec^{-2}}$, which is higher than that of the CLEAN image (\rev{480}$~{\rm mJy~arcsec^{-2}}$).

%====================================Fig====================================
\begin{figure*}[t]
\centering
\includegraphics[width=0.9 \textwidth]{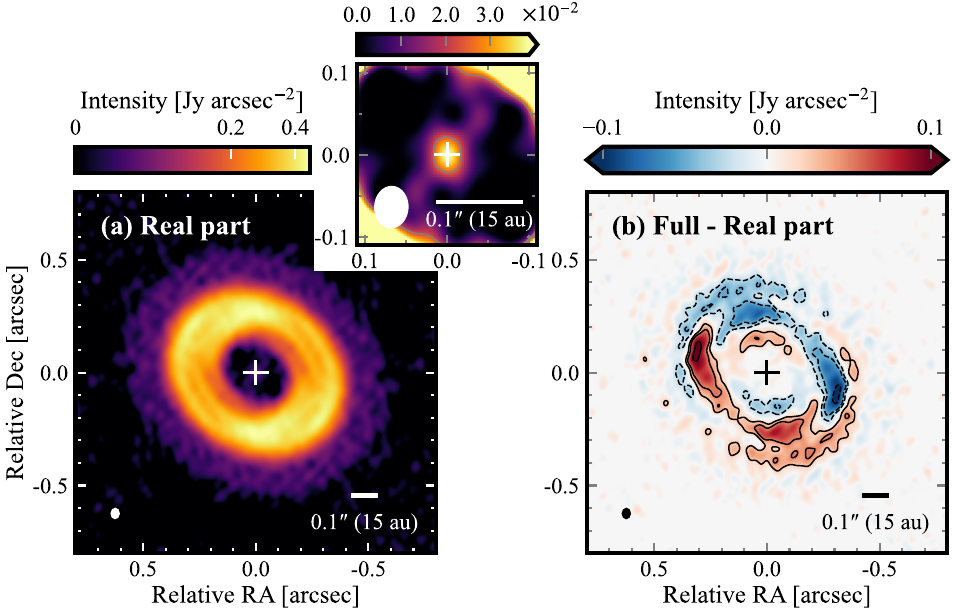}
\caption{
Real-part PRIISM image and the difference between the full- and real-part PRIISM images.
(a) PRIISM image reconstructed using only the real part of the visibilities, displayed with a color scale with a power-law exponent of 0.5. 
The inset \rev{shows the zoom-in view of the central region of the real-part PRIISM image.}
(b) Image obtained by subtracting the real-part PRIISM image from the PRIISM image reconstructed from the full visibilities. Contours are drawn at -4, -2, -1, 1, 2, and 4 $I_{\rm DT}$, with negative contours shown as dashed lines.
}

\label{fig:priism0p8_realimag}
\end{figure*}
%====================================Fig====================================

\section{Results} 
\label{sec:results}

\subsection{The CLEAN and PRIISM images}
\label{sec:results_cleanpriism}

\rev{The CLEAN and PRIISM images are shown in Figure~\ref{fig:cleanpriism}. These images} show a central cavity and a bright ring at radii of $\sim20$ au to $\sim70$ au, hereafter referred to as the main ring. The main ring is not azimuthally uniform and shows two bright crescents located at PAs of $\sim70^\circ$ and $\sim190^\circ$. \rev{Outside the main ring, a faint disk-skirt structure extending to a radius of $\sim80$ au is seen in all images.}
\rev{In addition, a compact emission component around the central star, defined as the inner disk, is seen in all the images, although in the CLEAN image with a robust parameter of $-0.2$ its peak intensity is below the $5\sigma$ level ($\simeq4.5\sigma$).}
These structures are consistent with those reported in previous ALMA studies \citep{UbeiraGabellini2019,Yamaguchi2024,Curone2025,Zagaria2025}. 
% The patchy features in the disk-skirt as seen in the PRIISM image are likely artifacts specific to the sparse modeling technique \citep{Yamaguchi2024}. 

\rev{The compact inner disk is detected above the $I_{\rm DT}$ level in the PRIISM image.} A two-dimensional Gaussian fit yields a size of $0.057'' \times 0.047'' ~(\rm{PA = 172^\circ})$. As this size is comparable to the PRIISM spatial resolution, the central disk remains unresolved with an upper limit of 8 au on its size. 
\rev{The flux density measured from the pixels above $I_{\rm DT}$ is $0.03~\mathrm{mJy}$. Assuming that the emission is optically thin, the dust mass of the inner disk is estimated as $M_{\rm dust}=F_\nu d^2/\{\kappa_\nu B_\nu(T_{\rm dust})\}$, where $\nu$ is the frequency, $F_\nu$ is the flux density, $d$ is the distance, $\kappa_\nu$ is the dust opacity, and $B_\nu(T_{\rm dust})$ is the Planck function at the dust temperature $T_{\rm dust}$. At a representative frequency of $\nu=225$ GHz, we adopt $\kappa_\nu=2.25$ cm$^2$ g$^{-1}$ \citep{Beckwith1990}. Assuming $T_{\rm dust}=50$--$200$ K, the dust mass is estimated to be $\sim1.7\times10^{-3}$--$7.3\times10^{-3}~M_\oplus$.}

\rev{Even though the spatial resolution of the CLEAN image with the robust parameter of $-0.2$ is comparable to that of the PRIISM image, the image contrasts in the PRIISM image are sharper. The PRIISM image exhibits the crescent features at PA $\sim70^\circ$ and $\sim190^\circ$ more clearly. The radial intensity variations in the PRIISM image are also sharper than those in the CLEAN image, particularly evident at PA of $\sim120^\circ$ and $\sim300^\circ$.} These sharp features may be related to the three arcs reported in the high-pass-filtered CLEAN image of \citet{Zagaria2025}, suggesting that PRIISM recovers these substructures more directly in the original image. 

\rev{To assess the imaging fidelity of these different reconstructions,} the PRIISM and CLEAN component images are Fourier-transformed and sampled onto the observed $uv$ coverage. The reduced chi-square values between these reconstructed visibilities and the original observed visibilities were evaluated in bins of $30~\mathrm{k}\lambda$ in deprojected $uv$ distance, adopting $\mathrm{PA}=230.8^\circ$ and $i=32.9^\circ$, as determined in Section~\ref{sec:global_disk_geometry}. 
The reduced chi-square value of the PRIISM visibilities ($\sim 1.2$) is indeed \rev{lower than that of the CLEAN components ($\sim 4.3$ and $\sim 5.3$ for robust parameters of $0.5$ and $-0.2$, respectively), indicating higher fidelity for the PRIISM reconstruction.}

In the PRIISM image, the brightness temperature $T_{\rm B}$ ranges from $3.8$ K to $18.7$ K over the radial extent of the main ring from $20 \rm~au$ to $70 \rm~au$. On the other hand, the inferred $T_{\rm dust}$ of the main ring passively heated by the central source with $L_* = 9L_\odot$ \citep{Wallack2024} is estimated to range from $25.8$ K to $48.3$ K, using Equation~(16) of \citet{Yamaguchi2024} with the flaring angle $\phi = 0.02$. A pixel-by-pixel comparison of the corresponding maps gives $T_{\rm dust}/T_{\rm B}$ ranging from $1.7$ to $12.7$. The 1.3 mm dust emission is thus most likely optically thin over most of the radial extent of the main ring.

\subsection{Real-part PRIISM Image}
\label{sec:results_realpart}

Because the real part of the visibilities traces the centro-symmetric component with respect to the phase center, we also reconstructed a PRIISM image using only the real part, allowing us to examine the underlying disk geometry centered on the stellar position while reducing the influence of non-centro-symmetric emission. For consistency with the PRIISM image reconstructed from the full visibilities, we used the same hyper-parameters, $(\Lambda_{l}, \Lambda_{\rm tsv}) = (3\times10^{5},~4\times10^{13})$, and adopted the same effective spatial resolution and detection threshold. The peak intensity is $\sim 460~{\rm mJy~arcsec^{-2}}$, corresponding to $\sim 86\%$ of that of the full PRIISM image. The bulk of the emission is thus contained in the real part. 

Figure~\ref{fig:priism0p8_realimag}(a) shows the real-part PRIISM image. Similar to the full PRIISM image, the real-part image shows the compact inner disk, central cavity, main ring, and faint extended disk-skirt structure. Near the inner edge of the main ring, a curved non-axisymmetric pattern of slightly reduced intensity is seen around PA $\sim130^\circ$. The same feature appears at PA $\sim310^\circ (=130^\circ+180^\circ)$ by construction of real-part imaging, which produces a centro-symmetric image. This pattern is further examined using high-pass filtering in Section~\ref{sec:analysis_highpass}.

The non-centro-symmetric component is highlighted in Figure~\ref{fig:priism0p8_realimag}(b), which was obtained by subtracting the real-part PRIISM image from the PRIISM image reconstructed from the full visibilities. Prominent positive residuals are seen to the south and east, spatially corresponding to the crescents in the full PRIISM image (Figure~\ref{fig:cleanpriism}). Additional residuals are distributed near the inner and outer edges of the main ring.

%====================================Fig====================================
\begin{figure*}[t]
\centering
\includegraphics[width=1.0\textwidth]{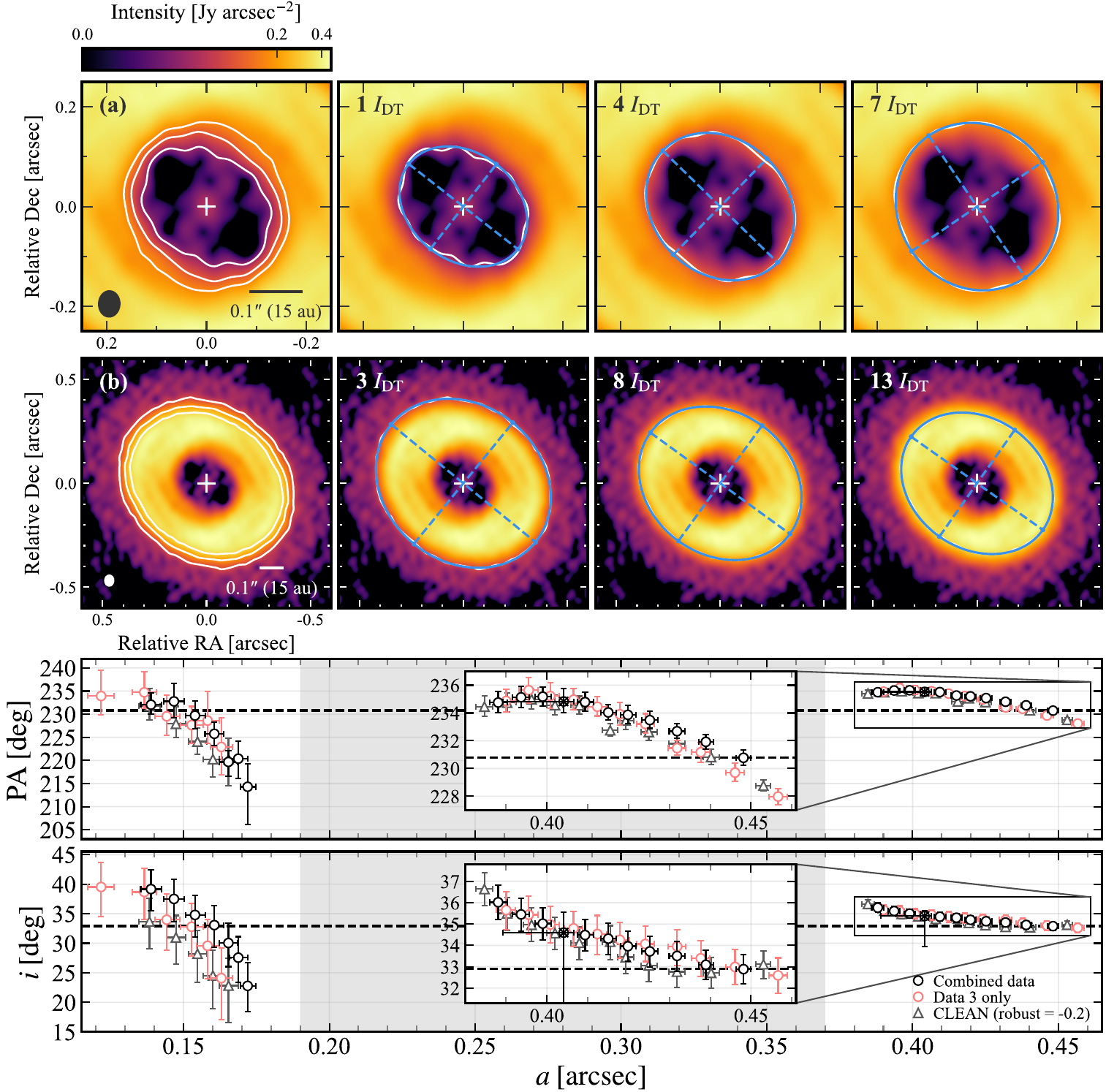}
\caption{
\rev{Ellipse-fitting results for the real-part images}. Ellipse fitting was performed on seven contours tracing the inner edge of the main ring (1--7 $I_{\rm DT}$) and eleven contours tracing the outer edge (3--13 $I_{\rm DT}$), in steps of $I_{\rm DT}$. The top and middle rows show the PRIISM image with a power-law stretch of exponent 0.3.
\textit{Top:}
White contours show three representative contours at 1, 4, and 7 $I_{\rm DT}$ tracing the inner edge. The panels to the right show the individual contours from panel (a), with the best-fit ellipses and their major and minor axes overlaid as blue solid curves and dashed lines, respectively.
\textit{Middle:}
Same as \rev{the top panels}, but for 3, 8, and 13 $I_{\rm DT}$ contours tracing the outer edge.
\textit{Bottom:}
\rev{PA and $i$ as functions of $a$ derived from ellipse fitting to the contours, with dashed lines indicating PA$_0$ and $i_0$. Each point represents an individual contour level. 
Black and red circles show the results for the real-part PRIISM images reconstructed from the combined data and Data 3 alone, respectively, and gray triangles show those for the real-part CLEAN image reconstructed from the combined data (robust = $-0.2$). The crossed black point near $a \sim 0.4''$ corresponds to the unstable fit to the 10 $I_{\rm DT}$ contour in the PRIISM image. The gray shaded region indicates the approximate radial range of the bright portion of the main ring excluded from the ellipse fitting.}}
\label{fig:ellipsefit_bfdpj_centerfixed_realonly}
\end{figure*}
%====================================Fig====================================

%====================================Fig====================================
\begin{figure*}[t]
\centering
\includegraphics[width=1.0\textwidth]{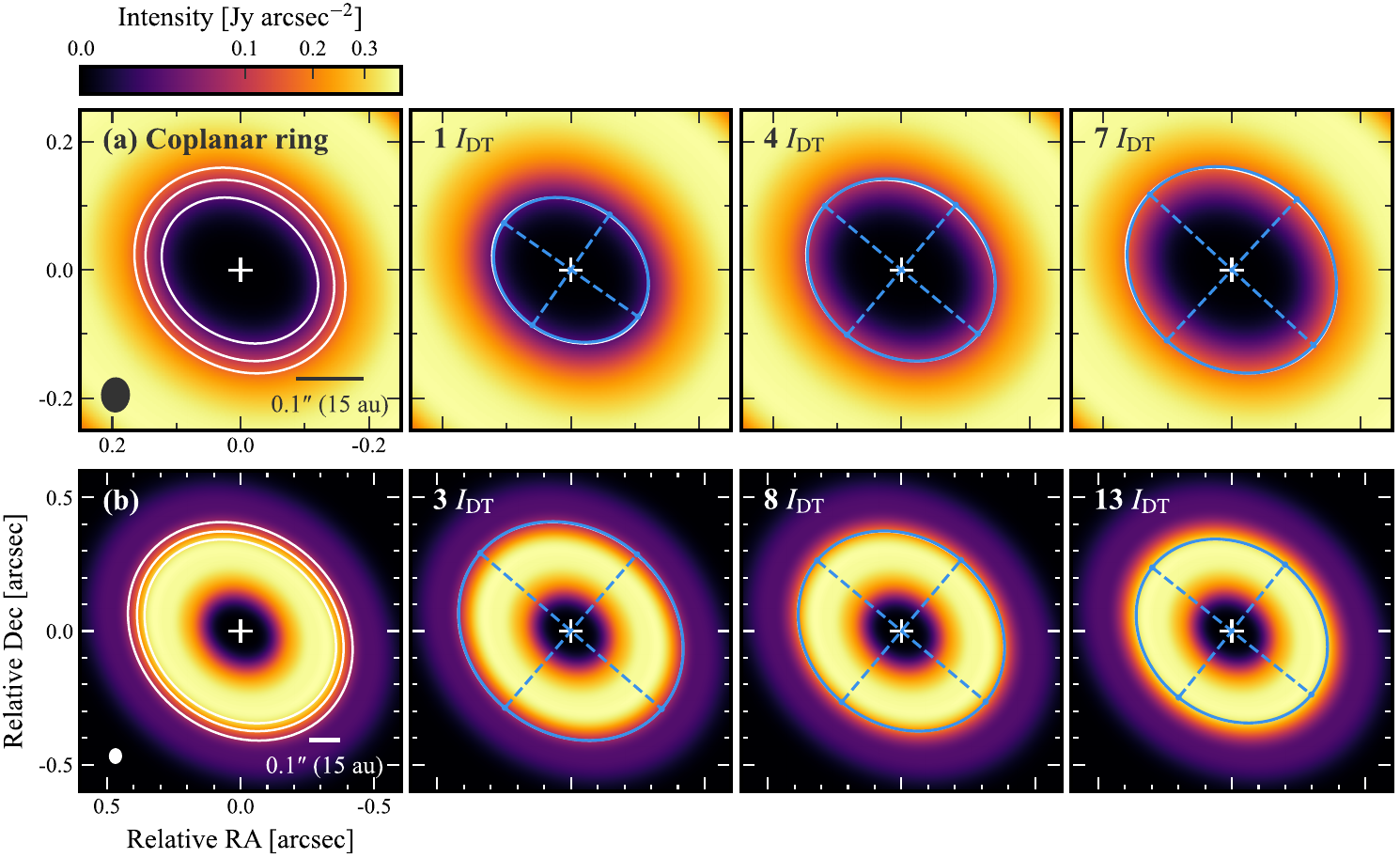}
\caption{ 
Coplanar ring model and its fitted ellipses. In each row, the left panel shows the model image with representative contours, and the panels to the right show individual contours with the best-fit ellipses and their major and minor axes in blue. The inner edge is traced in panel (a), and the outer edge is traced in panel (b).
}
\label{fig:gallery_flat}
\end{figure*}
%====================================Fig====================================

\section{Analysis} 
\label{sec:analysis}

\subsection{Disk Geometry from Contour Ellipse Fitting}
\label{sec:disk_geometry}

\subsubsection{Global Disk Geometry}
\label{sec:global_disk_geometry}

Throughout this paper, PA is measured east of north on the sky and ranges from $0^\circ$ to $360^\circ$. The inclination $i$ is defined as the angle between the disk plane and the plane of the sky, ranging from $0^\circ$ to $90^\circ$, with $i=0^\circ$ corresponding to a face-on disk. We denote the global major-axis PA and $i$ of the CQ~Tau dust disk by PA$_0$ and $i_0$, respectively. PA$_0$ is defined such that the far side of the disk lies along the minor axis at PA$_0+90^\circ$. 

\citet{Wolfer2021} interpreted the slightly brighter CO peak intensity on the northwestern side as evidence that this side is the far side. They attributed this brightness asymmetry to a larger projected area of the far-side CO-emitting surface and less severe beam dilution, arising from the vertical structure of the emitting layer. This orientation is further supported by the consistency between the observed spiral arm morphology and the disk rotation direction \citep{Wolfer2021,Hammond2022}, following the geometric interpretation presented in \citet{Muto2012}. We therefore adopt the northwestern side of the disk as the far side.

PA$_0$ and $i_0$ of the CQ~Tau dust disk were determined by performing ellipse fitting to a contour of the real-part PRIISM image. The ellipse center was fixed at the phase center. This choice reduces possible biases from non-centro-symmetric emission components. The free parameters are the lengths of the semimajor and semiminor axes ($a$, $b$) and the PA of the ellipse major axis. We assume that the continuum intensity approximately traces the dust density distribution since the 1.3 mm continuum emission is optically thin (Section~\ref{sec:results_cleanpriism}). The inclination is estimated as $i=\arccos(b/a)$. The detailed ellipse-fitting procedure, including the calculation of the fitting uncertainties and the reduced chi-square, $\chi^2_{\rm red}$, is described in Appendix~\ref{appendix:ellipsefit}.

An ellipse fit to the 3 $I_{\mathrm{DT}}$ contour tracing the outer edge of the main ring provides a reliable fit with $\chi_{\rm red}^2 \simeq 0.3$, yielding $\mathrm{PA} = 230.8^\circ \pm 0.6^\circ$ and $i = 32.9^\circ \pm 0.7^\circ$. These values are different from previous CO-based gas-disk orientations of $\mathrm{PA}\simeq235^\circ$ and $i\simeq35^\circ$--$36^\circ$ \citep{UbeiraGabellini2019,Izquierdo2025}. The CO-based values represent the global gas-disk orientation traced by line emission and kinematics, with the CO disk extending to a radius of $\sim150$ au \citep{UbeiraGabellini2019}, approximately twice the radial extent of the main dust ring. In contrast, our values are derived from the outer edge of the dust ring. Since this paper focuses on the dust disk structure, we adopt the continuum-based values, $\mathrm{PA}_0=230.8^\circ$ and $i_0=32.9^\circ$, throughout this paper.
%====================================Fig====================================
\begin{figure*}[t]
\centering
\includegraphics[width=1.0\textwidth]{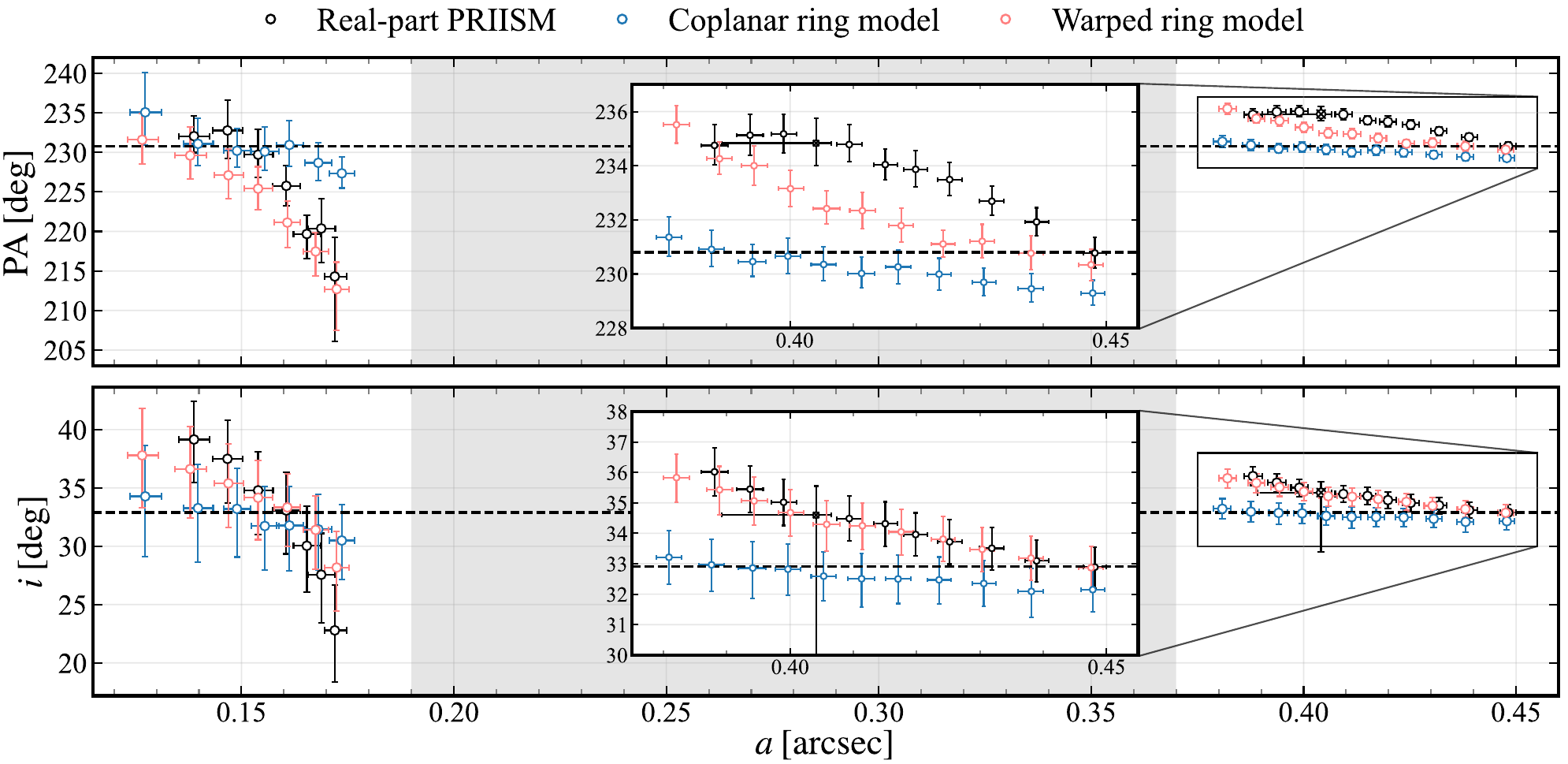}
\caption{
Comparison of the ellipse-fitting results for the real-part PRIISM image and the toy models. The upper and lower panels show PA and $i$ as functions of $a$, respectively. Black, blue, and red markers indicate the results for the real-part PRIISM image, the coplanar ring model, and the warped ring model, respectively. The horizontal dashed lines indicate the global disk orientation, PA$_0$ and $i_0$. The gray shaded region is the same as that in Figure~\ref{fig:ellipsefit_bfdpj_centerfixed_realonly}.
}
\label{fig:ellipsefit_warpmodel}
\end{figure*}
%====================================Fig====================================

\subsubsection{Radial Variations in Disk Geometry}
\label{sec:radial_variations_geometry}

The same ellipse-fitting analysis \rev{is} applied over the ranges of contour levels tracing the inner and outer edges of the main ring, from 1 $I_{\rm DT}$ to 7 $I_{\rm DT}$ and from 3 $I_{\rm DT}$ to 13 $I_{\rm DT}$, respectively, in steps of $I_{\rm DT}$. The fitting results are shown in Figure~\ref{fig:ellipsefit_bfdpj_centerfixed_realonly}, which summarizes the resulting PA and $i$ as functions of $a$. Here, $a$ is used as a proxy for the disk radius at each contour level. At both the inner and outer edges, the PA and $i$ vary with $a$ beyond the fitting uncertainties. At the inner edge, over $a \sim 0.14''$--$0.17''$ ($21$--$26$ au), the PA and $i$ span approximately $214^\circ$--$233^\circ$ and $23^\circ$--$39^\circ$, respectively. At the outer edge, over $a \sim 0.39''$--$0.45''$ ($58$--$67$ au), they span approximately $231^\circ$--$235^\circ$ and $33^\circ$--$36^\circ$, respectively. \rev{The gray shaded region in Figure \ref{fig:ellipsefit_bfdpj_centerfixed_realonly} approximately corresponds to the radial range of the emission ridge of the main ring, where the contours are affected by azimuthal intensity variations and do not simply trace the underlying disk geometry. This region is thus excluded from the ellipse-fitting analysis, which aims to reveal radial variations in the disk orientation. At the inner edge of the main ring,} the 8 $I_{\rm DT}$ contour is locally distorted by the curved reduced-intensity pattern shown in Figure~\ref{fig:priism0p8_realimag}(a). At the outer edge, the 14 $I_{\rm DT}$ contour traces only the brighter azimuthal portions of the ring rather than the overall outer edge.

\rev{To examine whether the radial variations in PA and $i$ depend on the imaging method, the same ellipse fitting procedure is applied to the CLEAN image reconstructed from the real part of the visibilities with the robust parameter of $-0.2$ at the same intensity levels as for the real-part PRIISM image. The results from the 1 $I_{\rm DT}$ contour are excluded because its intensity is below the $5\sigma$ level in the CLEAN image ($\sim3\sigma$), and that of the 7 $I_{\rm DT}$ contour tracing the inner edge is discarded because of the fitting instability. The remaining data points show radial trends largely consistent with those obtained from the real-part PRIISM image at both the inner and outer edges (Figure~\ref{fig:ellipsefit_bfdpj_centerfixed_realonly}), indicating that the radial variations are not specific to the PRIISM reconstruction.}

To assess a possible effect of image smearing due to the combination of multi-epoch data, \rev{the ellipse fitting is also applied to the real-part PRIISM image reconstructed from Data 3 alone, adopting $(\Lambda_l,~\Lambda_{\rm tsv})=(2\times10^5,~7\times10^{13})$ (Section~\ref{sec:imaging_with_CLEAN_and_PRIISM}). The fitting is conducted at the same intensity levels as that of the combined-data real-part PRIISM image, but the 7 $I_{\rm DT}$ contour tracing the inner edge is excluded because the fitting is unstable. The overall trends are again} broadly consistent with those obtained from the combined data \rev{(Figure~\ref{fig:ellipsefit_bfdpj_centerfixed_realonly})}, indicating that the radial variations in PA and $i$ are not primarily driven by multi-epoch smearing.

We further \rev{test} a simple eccentric-orbit interpretation by applying center-free ellipse fitting to the full PRIISM image at the same contour levels as in the center-fixed analysis. In an eccentric orbit, the central star is expected to lie at one focus of the ellipse \citep{Dong2018,Kuo2022}. However, the fitted ellipse centers \rev{are} typically located within one third of the effective spatial resolution from the phase center, whereas, for every contour, the distance from the phase center to the nearer focus \rev{is} more than three times \rev{the distance} to the ellipse center. The observed geometry therefore disfavors a simple eccentric-orbit interpretation.

\subsection{Toy Modeling}
\label{sec:analysis_toymodels}

\subsubsection{Coplanar Ring Model}
\label{sec:analysis_basetoy}

The base disk structure is derived from the fitting to the real-part PRIISM image of the main ring with a coplanar ring model of multi-Gaussian functions. The model disk is assumed to be infinitesimally thin. Appendix~\ref{appendix:make_basetoy} describes the detailed construction of the coplanar ring model. The model image was convolved to the same resolution as the PRIISM image and analyzed using the same center-fixed ellipse-fitting procedure and contour levels as for the observations.

Figures~\ref{fig:gallery_flat} and \ref{fig:ellipsefit_warpmodel} show the coplanar ring model images and their ellipse-fitting results. At both the inner and outer edges, the PA and $i$ profiles exhibit only modest radial variations and remain close to PA$_0$ and $i_0$, failing to reproduce the observed variations, particularly around $a \simeq 0.17''$. To assess the modest variations in PA and $i$, we repeated the ellipse fitting on a coplanar model convolved with a circular Gaussian whose FWHM was set to the geometric-mean resolution. The variations were largely reduced but did not disappear completely. Thus, the variations likely reflect not only the elliptical resolution but also uncertainties associated with extracting and fitting the pixelized contours.

\subsubsection{Warped Ring Model}
\label{sec:analysis_warped_ring}

%====================================Fig====================================
\begin{figure}[t]
\centering
\includegraphics[width=0.4 \textwidth]{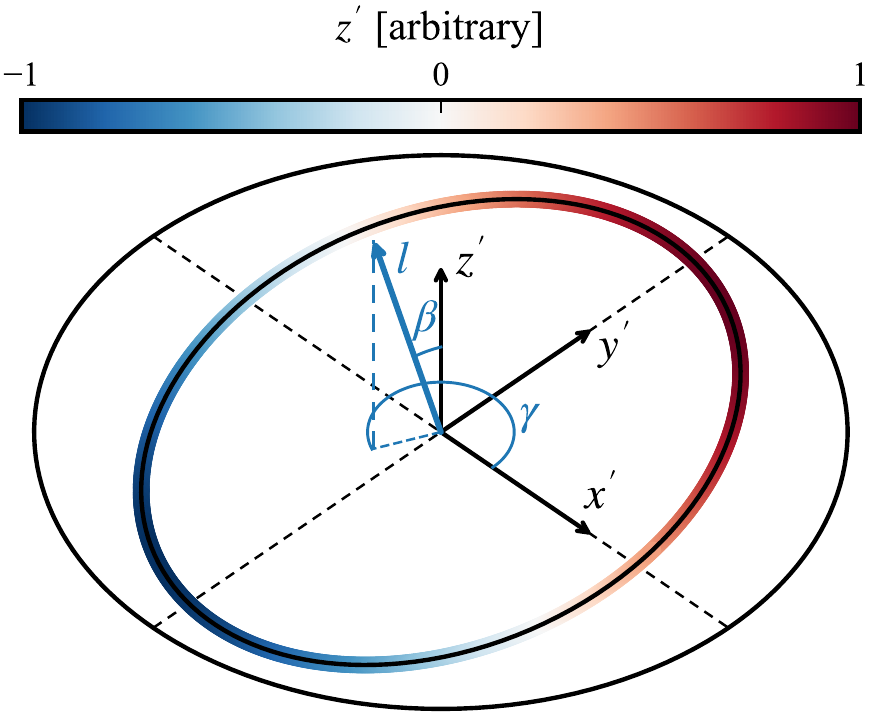}
\caption{
Schematic illustration of the coordinate system adopted to characterize the warp geometry, based on \citet{Juhasz2017}. The black arrows indicate the global disk coordinate system $(x', y', z')$, where the $(x', y')$ plane and $z'$ axis correspond to the global midplane and global disk normal defined by PA$_0$ and $i_0$, respectively. The colored annulus illustrates the warp geometry. Its local disk-normal vector is denoted by $l$. $\beta(r)$ represents the local tilt angle from the global disk normal, while $\gamma(r)$ represents the azimuthal direction of the tilt.
}
\label{fig:schematic_betagamma}
\end{figure}
%====================================Fig====================================

A warped ring model was constructed by applying a radial profile of the disk-normal direction to the coplanar ring model. The disk-normal direction at each radius (i.e., warp) is described by the polar angle $\beta(r)$ and azimuthal angle $\gamma(r)$, as illustrated in Figure~\ref{fig:schematic_betagamma}. $\beta$ is defined over $0^\circ \leq \beta \leq 90^\circ$. $\gamma$ is taken to lie in the range $-180^\circ < \gamma \leq 180^\circ$, with $\gamma=0^\circ$ corresponding to the direction along the global disk major axis at PA$_0$.

The warp geometry is represented by a set of thin annuli. For each annulus, the emission \rev{is} projected onto the sky plane according to the local values of $\beta(r)$ and $\gamma(r)$ while conserving its flux, as appropriate for the observed optically thin continuum emission. As a result, when an annulus is projected onto a smaller area on the sky plane, its intensity can be locally enhanced.

We modeled the radial profiles of $\beta$ and $\gamma$ as continuous piecewise-linear functions with a common break radius $r_{\rm ref}$:
\begin{align}
\beta(r) &=
\begin{cases}
\beta_0+s_{\beta,\mathrm{inner}}r,
& r \leq r_{\rm ref}, \\[2pt]
\beta_0+s_{\beta,\mathrm{inner}}r_{\rm ref}
+s_{\beta,\mathrm{outer}}(r-r_{\rm ref}),
& r>r_{\rm ref},
\end{cases}
\\
\gamma(r) &=
\begin{cases}
\gamma_0+s_{\gamma,\mathrm{inner}}r,
& r \leq r_{\rm ref}, \\[2pt]
\gamma_0+s_{\gamma,\mathrm{inner}}r_{\rm ref}
+s_{\gamma,\mathrm{outer}}(r-r_{\rm ref}),
& r>r_{\rm ref}.
\end{cases}
\end{align}
If the piecewise-linear expression for $\beta(r)$ becomes negative, we set $\beta(r)=0^\circ$. $r_{\rm ref}$ was fixed at $0.3''$, approximately the radial midpoint of the main ring. $\beta_0$ and $\gamma_0$ are the values at $r=0$, while $s_{\beta,\mathrm{inner}}$ and $s_{\beta,\mathrm{outer}}$ are the radial slopes of $\beta(r)$ inside and outside the break radius, respectively; $s_{\gamma,\mathrm{inner}}$ and $s_{\gamma,\mathrm{outer}}$ are defined analogously for $\gamma(r)$.

\begin{figure}[t]
\centering
\includegraphics[width=0.47\textwidth]{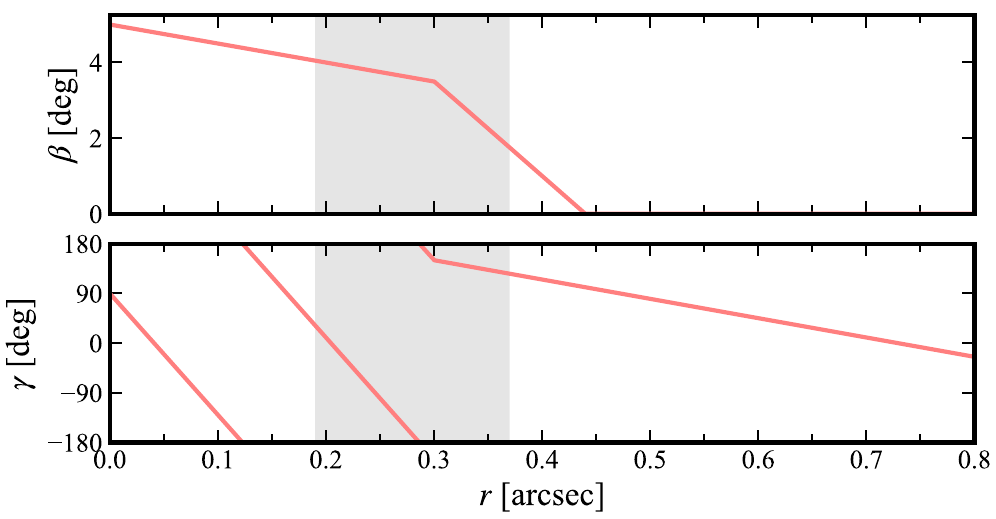}
\caption{ 
\rev{Fiducial $\beta(r)$ and $\gamma(r)$ profiles. The gray shaded region indicates the radial range of the main dust ring excluded from the ellipse fitting, as also shown in Figure~\ref{fig:ellipsefit_bfdpj_centerfixed_realonly}.}
}
\label{fig:warpprof_input}
\end{figure}

\begin{figure*}[t]
\centering
\includegraphics[width=1.0 \textwidth]{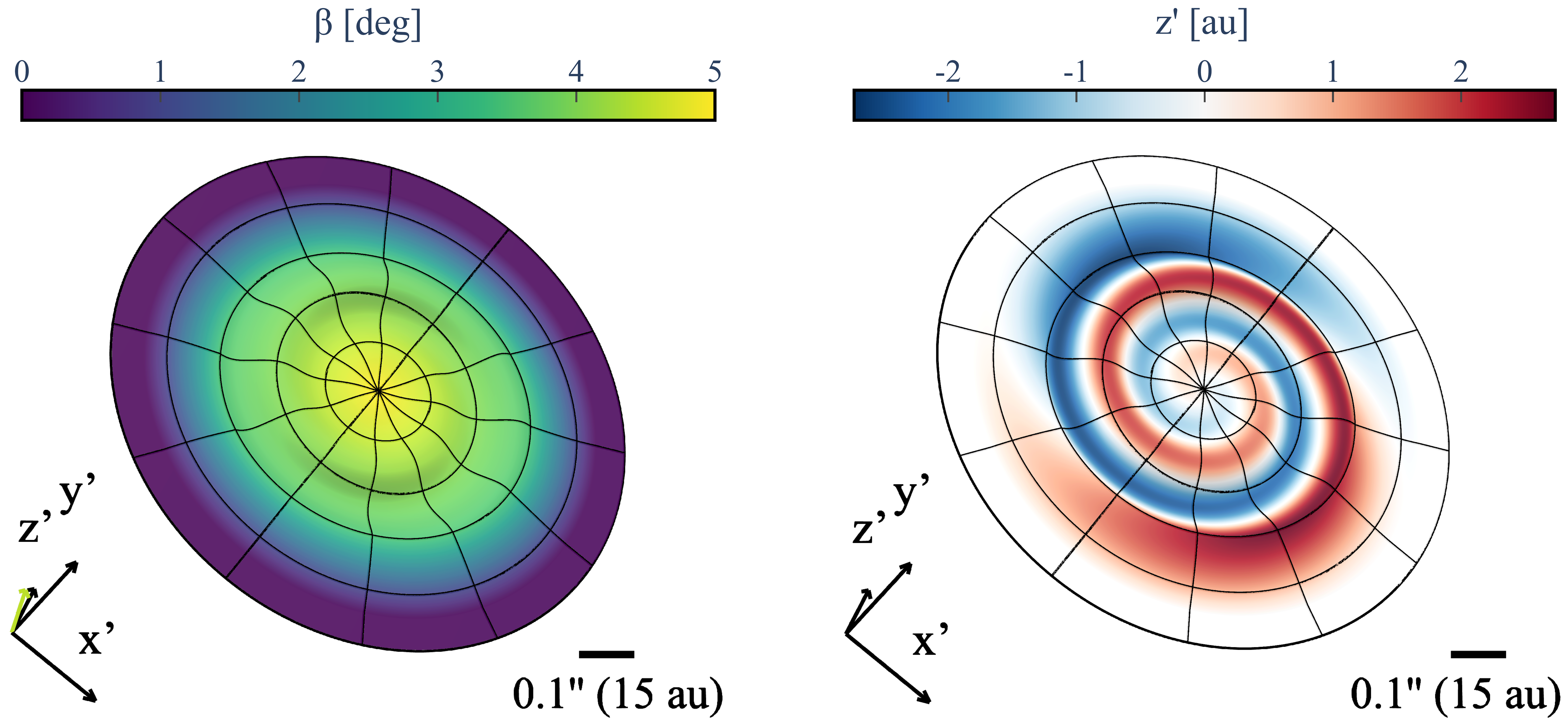}
\caption{
The warp geometry of the fiducial warped ring model. The black arrows indicate the global disk coordinate system $(x', y', z')$ defined by PA$_0$ and $i_0$. 
\textit{Left:} The warp geometry colored by $\beta(r)$. The colored arrow in the lower-left corner indicates the local disk-normal vector of the annulus at $r = 0.1''$.
\textit{Right:} Same as the left panel, but colored by $z'$, the height of the warped-disk midplane relative to the global midplane, $(x', y')$. The radial and azimuthal grid lines are drawn every $0.1''$ and $30^\circ$, respectively, in the disk coordinates.
}
\label{fig:schematic_warpmodel}
\end{figure*}

We manually explored the parameters defining the $\beta(r)$ and $\gamma(r)$ profiles. Each trial model was convolved to the same resolution as the PRIISM image and analyzed using the same center-fixed ellipse-fitting procedure and contour levels as for the observations. The fiducial model was selected so that its ellipse-fitting results broadly reproduce the observed radial variations in PA and $i$. The explored parameter ranges and fiducial values are listed in Table~\ref{table:warp_params}, and the resulting fiducial $\beta(r)$ and $\gamma(r)$ are shown in Figure~\ref{fig:warpprof_input}. In the fiducial model, $\beta$ decreases outward and reaches $0^\circ$ at $r\simeq0.44''$, while $\gamma$ varies with radius. The resulting warp geometry is shown in Figure~\ref{fig:schematic_warpmodel}.

\begin{figure*}[t]
\centering
\includegraphics[width=1.0\textwidth]{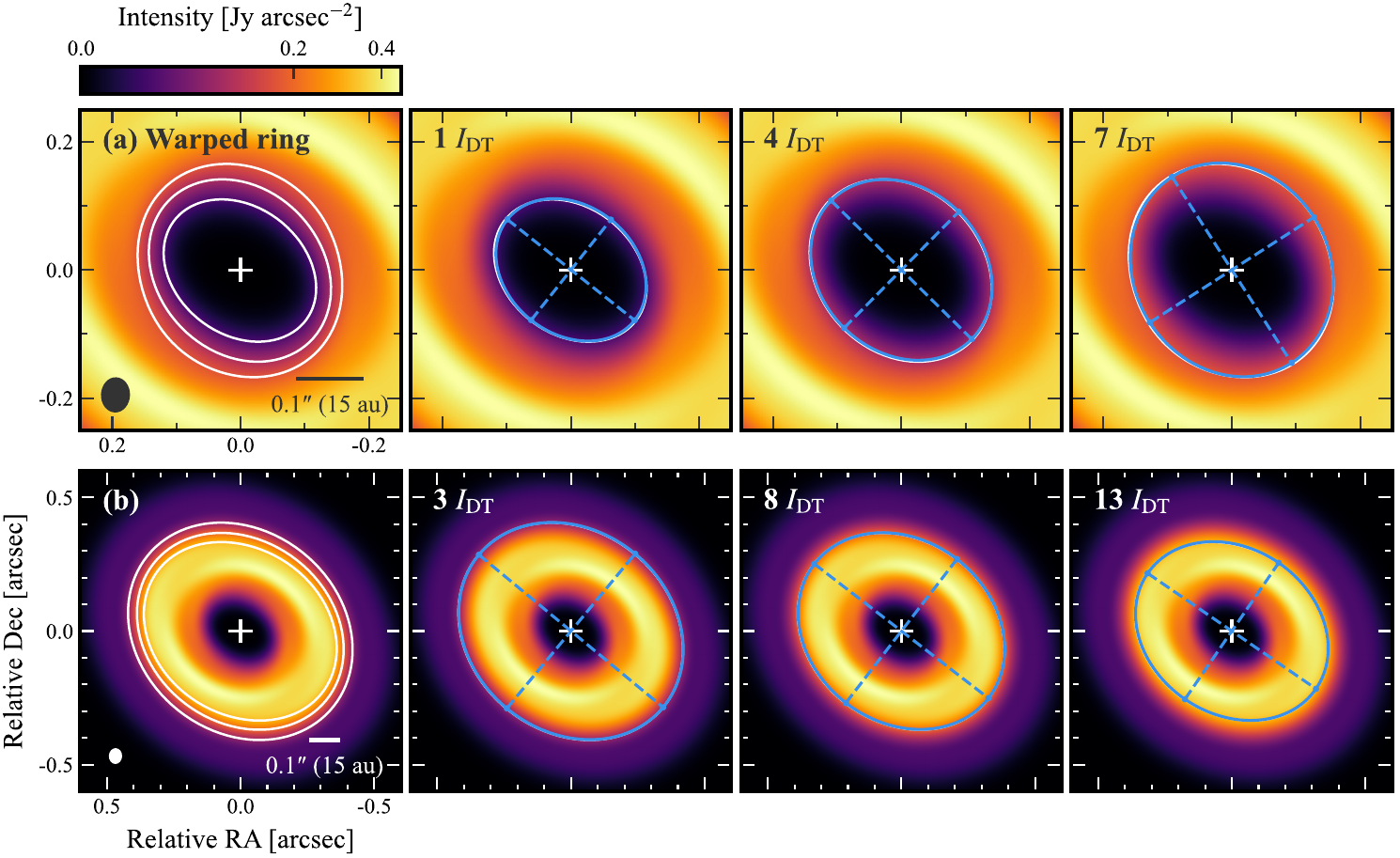}
\caption{ 
Same as Figure~\ref{fig:gallery_flat}, but for the warped ring model.
}
\label{fig:gallery_warp}
\end{figure*}

Figures~\ref{fig:ellipsefit_warpmodel} and \ref{fig:gallery_warp} show the ellipse-fitting results and the warped ring model images, respectively. At the inner edge, the model PA decreases by $\sim20^\circ$, from $\sim232^\circ$ to $\sim213^\circ$. The model $i$ decreases by $\sim10^\circ$, from $\sim38^\circ$ to $\sim28^\circ$. At the outer edge, the model PA decreases by $\sim5^\circ$, from $\sim236^\circ$ to $\sim230^\circ$, while the model $i$ decreases by $\sim3^\circ$, from $\sim36^\circ$ to $\sim33^\circ$. The model (red markers in Figure~\ref{fig:ellipsefit_warpmodel}) broadly reproduces the observed trends and amplitudes of the variations (black markers). 
\rev{At the inner edge, the differences between the model and observed PA and $i$ are comparable to the typical ellipse-fitting uncertainties ($\sim 4^\circ$), indicating that the model reasonably reproduces the observed variations. At the outer edge, a systematic} discrepancy of $\lesssim 3^\circ$ remains between the model and observed PA profiles, possibly due to the simplified parametric warp profile or unmodeled intensity structures. By contrast, the coplanar model does not reproduce the observed variations (blue markers), suggesting that the warped ring model is a better representation of the observed image.

%-----------------------------Table Start-----------------------------
\begin{table}[t]
\centering
\refstepcounter{table}
{\centering \textbf{Table \thetable}\par}
{\centering Explored parameter ranges and fiducial values for the warped ring model\par}
\label{table:warp_params}
\vspace{3pt}
\footnotesize
\renewcommand{\arraystretch}{1.15}
\setlength{\tabcolsep}{0pt}

\begin{tabular*}{\columnwidth}{
    @{\extracolsep{\fill}}
    l
    c
    c
    @{}
}
\hline\hline
\noalign{\vskip 2pt}
Parameter
& \shortstack{Explored\\range}
& \shortstack{Fiducial\\value} \\
\hline

$\beta_0$ [deg]
& $1$~--~$30$
& $5$ \\

$s_{\beta,\mathrm{inner}}$ [deg arcsec$^{-1}$]
& $-90$~--~$-2$
& $-5$ \\

$s_{\beta,\mathrm{outer}}$ [deg arcsec$^{-1}$]
& $-40$~--~$-5$
& $-25$ \\

$\gamma_0$ [deg]
& $-180$~--~$180$
& $90$ \\

$s_{\gamma,\mathrm{inner}}$ [deg arcsec$^{-1}$]
& $-2400$~--~$-2000$
& $-2200$ \\

$s_{\gamma,\mathrm{outer}}$ [deg arcsec$^{-1}$]
& $-1000$~--~$-200$
& $-350$ \\

\noalign{\vskip 1pt}
\hline
\end{tabular*}

\vspace{3pt}
\tablecomments{\parbox{\columnwidth}{
\textbf{Note.}
The explored ranges indicate the minimum and maximum values considered. A total of 384 trial models were explored.
}}
\end{table}
%------------------------------Table End------------------------------

% ====================================Fig====================================
%====================================Fig====================================

%====================================Fig====================================
%====================================Fig====================================

%====================================Fig====================================
%====================================Fig====================================

\subsubsection{Comparison of High-pass-filtered Images}
\label{sec:analysis_highpass}

Figure~\ref{fig:gallery_HpFwhm0p1_main} compares the high-pass-filtered real-part PRIISM image with those of the coplanar and warped ring models. Each image was obtained by subtracting its Gaussian-smoothed version with an FWHM of $0.1''$ from the original image. The high-pass-filtered real-part PRIISM image shows multiple curved residual patterns distributed in a centro-symmetric manner. 
In particular, curved negative residuals along the inner edge and narrow positive residuals near the outermost radius are seen, delineated by the white and black dashed lines, respectively. The former correspond to the curved pattern of reduced intensity seen at PAs of $\sim130^\circ$ and $\sim310^\circ$ in the original real-part PRIISM image (Figure~\ref{fig:priism0p8_realimag}(a)). By contrast, the coplanar ring model shows only axisymmetric residuals associated with the multi-Gaussian ring components. In the warped model, annuli at different radii have different sky-plane projections, producing radial and azimuthal intensity variations. Such behavior appears consistent with that in the high-pass-filtered PRIISM image, while the detailed emission distributions do not perfectly match. 
The high-pass-filtered image of the warped model indeed reproduces both the curved negative residuals along the inner edge and the narrow positive residuals near the outermost radius (Figure~\ref{fig:gallery_HpFwhm0p1_main}), similar to those in the real-part PRIISM image. These qualitative trends persist for the different smoothing FWHMs examined in Appendix~\ref{appendix:highpass}.

\section{Discussion}
\label{sec:discussion}

\subsection{Evidence for the Dust Warp}
\label{sec:overall_disk_structure}

Our PRIISM imaging of the real-part visibilities enables us to investigate the detailed, systematic disk structures while reducing the influence of non-centro-symmetric components. 
While the coplanar ring model fails to reproduce the observed radial variations in both PA and $i$, the fiducial warped ring model, in which $\beta$ decreases from $\sim4^\circ$ to $1^\circ$ for $r=20$ to 60 au, provides a better match to these variations. Although this toy model is based on simple parametric warp profiles and is not intended to provide a unique quantitative fit, the comparison suggests that the main dust ring in CQ~Tau is warped. Further support for the dust warp comes from the center-free ellipse-fitting results, which disfavor a simple eccentric-orbit interpretation (Section~\ref{sec:radial_variations_geometry}), and from the agreement between the observed high-pass-filtered morphology and that of the warped ring model (Section~\ref{sec:analysis_highpass}). The latter comparison also suggests that high-pass filtering may provide a supplementary tool for identifying disk warps.

A warped disk geometry is intrinsically three-dimensional, because the disk height relative to the global disk midplane varies with radius and azimuth. In addition to the ring-scale warp that gradually weakens toward the outer edge, CQ~Tau hosts a strongly misaligned inner disk with a tilt of $\beta \sim 40^\circ$ \citep{Bohn2022}. Furthermore, \citet{Winter2025} showed that the non-axisymmetric CO line-of-sight velocity pattern over radii of $\sim45$--$155$ au can be reproduced by the velocity field expected for a warped disk with a tilt of a few degrees. The CQ~Tau disk may therefore be interpreted as a hierarchical three-dimensional disk structure.

%====================================Fig====================================
\begin{figure*}[t]
\centering
\includegraphics[width=1.0\textwidth]{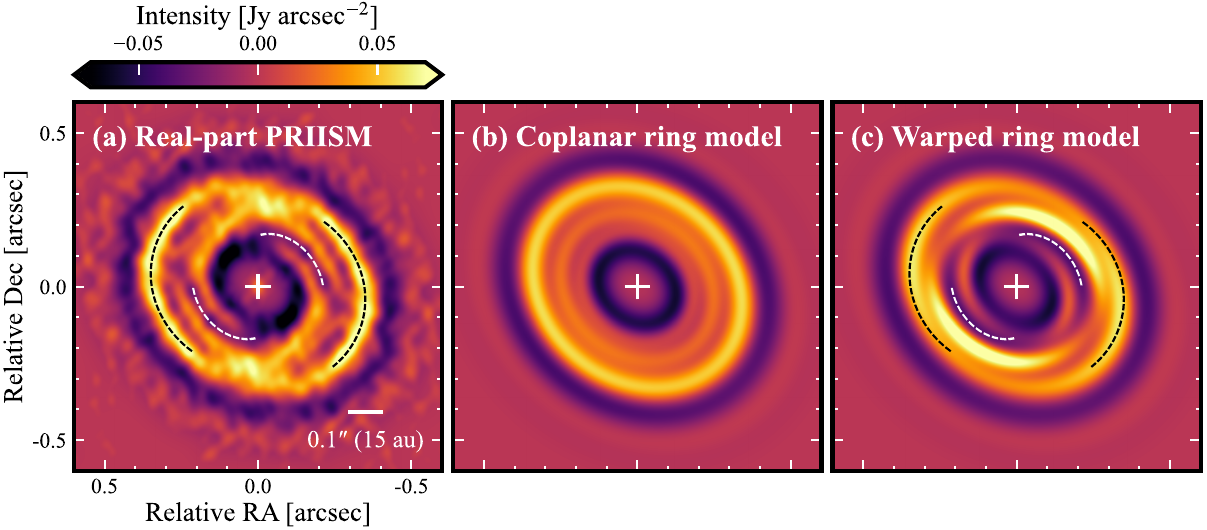}
\caption{
High-pass-filtered images of the real-part PRIISM image, the coplanar ring model, and the warped ring model. Each image was obtained by subtracting its Gaussian-smoothed version with an FWHM of $0.1''$. The curved negative residuals along the inner edge and the narrow positive residuals near the outermost radius in panels (a) and (c) are delineated by the white and black dashed lines, respectively.
}
\label{fig:gallery_HpFwhm0p1_main}
\end{figure*}
%====================================Fig====================================

\subsection{Physical Origin of the Warp}
\label{sec:warporigins}

\subsubsection{Companion}
\label{sec:warporigin_companion}
A companion on an inclined orbit exerts a secular gravitational torque on the disk and induces nodal precession of the disk angular momentum vector \citep{Juhasz2017,Facchini2018}. Since the precession rate generally depends on radius, different annuli precess at different rates if radial communication within the disk is inefficient, potentially leading to disk breaking. In the bending-wave regime ($\alpha < h/r$), where $\alpha$ is the viscosity parameter in the $\alpha$ prescription \citep{Shakura1973} and $h/r$ is the disk aspect ratio, such differential precession can be communicated across the disk by bending waves. The radial communication timescale, $t_{\rm c}$, and the global nodal precession timescale, $t_{\rm n,global}$, therefore characterize how the disk responds to the companion torque \citep[e.g.,][]{Lubow2018,Smallwood2019,Bi2020}. When $t_{\rm c}/t_{\rm n,global}>1$, differential precession proceeds faster than radial communication, and the disk can break into distinct regions. In contrast, when $t_{\rm c}/t_{\rm n,global}<1$, radial communication is faster than global precession, so the warped disk responds as a single coherent unit.

Following \citet{Bi2020}, we estimate $t_{\rm c}$ and $t_{\rm n,global}$ for the main ring of CQ~Tau, assuming that the ring is in the bending-wave regime. For the disk aspect ratio, we adopt $h/r=0.125(r/56~{\rm au})^{0.05}$, based on the radiative-transfer modeling of \citet{UbeiraGabellini2019}. For a companion on a circular orbit, these timescales are given by
\begin{equation}
t_{\rm c} \simeq
\frac{8}{5\Omega_{\rm comp}(h/r)_{\rm out}}
\left(\frac{r_{\rm out}}{a_{\rm comp}}\right)^{3/2},
\end{equation}
\begin{equation}
t_{\rm n,global}
=
\frac{
8\pi\,(M_\ast+M_{\rm comp})^2
}{
3M_\ast M_{\rm comp}
\left\langle
\left(\frac{a_{\rm comp}}{r}\right)^{7/2}
\right\rangle
\Omega_{\rm comp}
},
\end{equation}
\begin{equation}
\left\langle
\left(\frac{a_{\rm comp}}{r}\right)^{7/2}
\right\rangle
=
\frac{
\displaystyle
\int_{r_{\rm in}}^{r_{\rm out}}
\Sigma(r) r^3 \Omega(r)
\left(\frac{a_{\rm comp}}{r}\right)^{7/2}
\,dr
}{
\displaystyle
\int_{r_{\rm in}}^{r_{\rm out}}
\Sigma(r) r^3 \Omega(r)
\,dr
},
\end{equation}
where $r_{\rm in}$ and $r_{\rm out}$ define the radial range of the main ring over which the global precession is evaluated. The quantity $(h/r)_{\rm out}$ is the disk aspect ratio at $r_{\rm out}$. The angle brackets denote the angular-momentum-weighted average over this radial range, where $\Sigma(r)$ and $\Omega(r)$ are the disk surface density and angular frequency at radius $r$, respectively. We assume $\Sigma(r)\propto r^{-3/2}$ in evaluating this average, following \cite{Bi2020}. $M_\ast$ is the stellar mass, and $M_{\rm comp}$ and $a_{\rm comp}$ are the companion's mass and orbital radius, respectively. $\Omega_{\rm comp}$ is the orbital angular frequency of the companion.
Here, we adopt $r_{\rm in}=20~{\rm au}$, $r_{\rm out}=70~{\rm au}$, $M_\ast=1.6~M_\odot$ \citep{Wallack2024}, and $(h/r)_{\rm out}=0.126$ \citep{UbeiraGabellini2019}. 
Because the companion has not been directly detected, we adopt the companion solution inferred from the Gaia DR2--DR3 proper-motion-anomaly analysis of \citet{Vioque2026}, $M_{\rm comp}=309~{\rm M_J}$ and $a_{\rm comp}=1~{\rm au}$, and assume a circular orbit. We do not assign a specific value to the companion--disk mutual inclination, because the following timescale comparison does not explicitly depend on this angle. This gives a radial communication timescale of $t_{\rm c}\sim 9 \times 10^{-4}$ Myr and a global precession timescale $t_{\rm n,global}\sim 1.7$ Myr, corresponding to $t_{\rm c}/t_{\rm n,global} \sim 5 \times 10^{-4}$. In this case, radial communication is efficient compared to global precession, so the main ring is expected to respond as a single structure, consistent with the main ring responding coherently as a warped structure. Because the proper-motion-anomaly analysis is subject to a broad mass--orbit degeneracy \citep{Vioque2026,Blakely2026}, we also explore an extended parameter range, with $M_{\rm comp}$ up to $1000~{\rm M_J}$ and $a_{\rm comp}$ up to $7$ au. The upper limit on $a_{\rm comp}$ corresponds to about one-third of the central cavity radius of CQ~Tau, $r_{\rm cav}\sim20$ au \citep{UbeiraGabellini2019}. This choice keeps the main ring at radii larger than several times the companion orbit, where the secular torque formulas provide a reasonable approximation \citep{Lubow2018,Smallwood2019}. Across this parameter space, $t_{\rm c}/t_{\rm n,global}$ remains below unity, indicating that the main ring is expected to respond as a single structure over a wide range of companion parameters, consistent with the warped-ring interpretation in Section~\ref{sec:overall_disk_structure}.

\begin{figure}[t]
\centering
\includegraphics[width=0.4\textwidth]{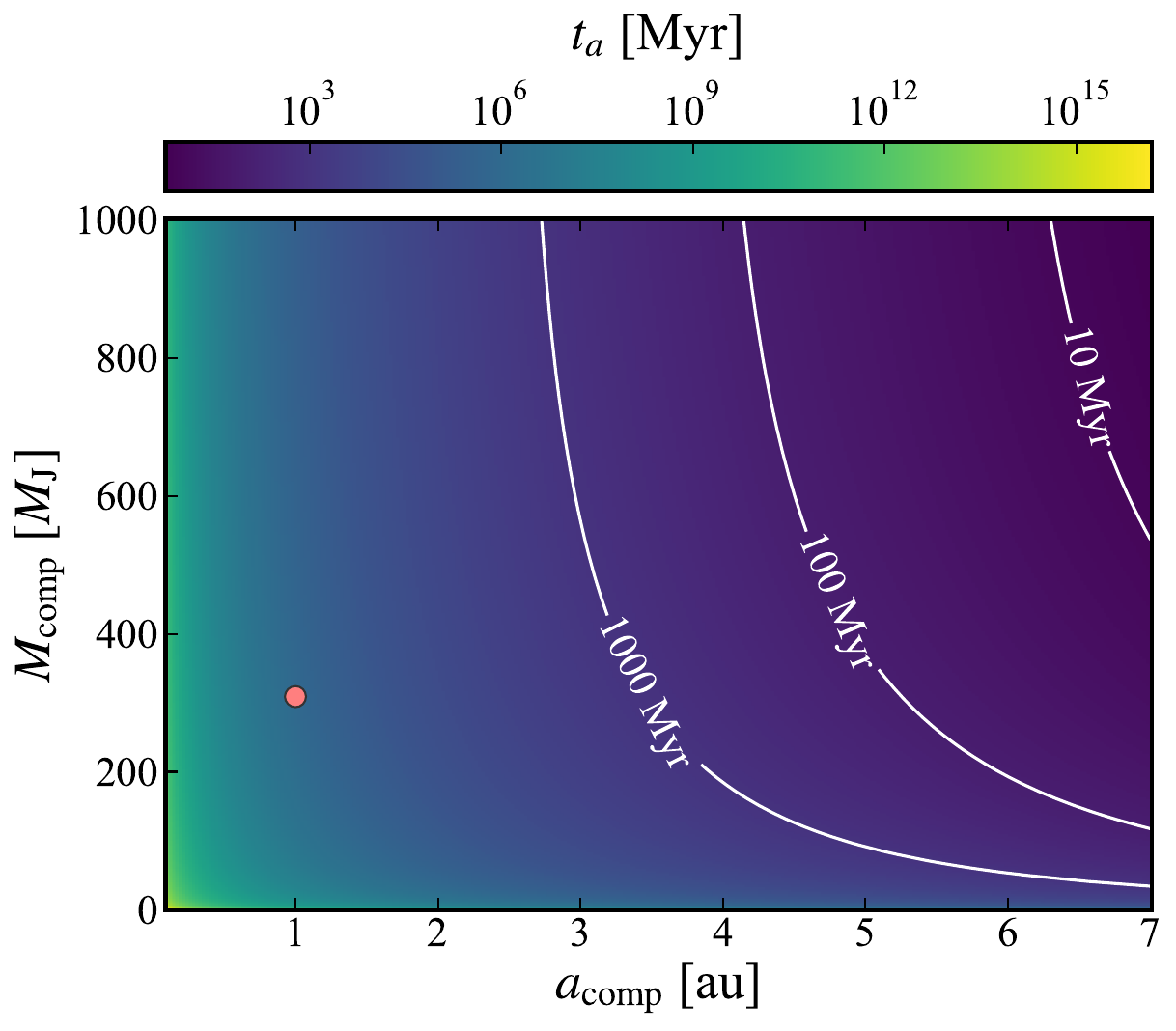}
\caption{
Alignment timescale $t_a$ as a function of companion semimajor axis $a_{\rm comp}$ 
and companion mass $M_{\rm comp}$ for $\alpha = 0.01$. 
The color scale shows $t_a$, and the solid contours indicate 
$t_a = 10$, 100, and 1000 Myr. 
For reference, the red point shows a set of companion parameters, $M_{\rm comp} = 309\,M_{\rm J}$ and $a_{\rm comp} = 1\,{\rm au}$, inferred by \citet{Vioque2026}.
}
\label{fig:ta_map}
\end{figure}

This scenario requires that the inclined disk--companion configuration can be maintained over a sufficiently long timescale against viscous alignment. Following \citet{Bi2020}, we estimate the alignment timescale, $t_{\rm a}$, over which the mutual inclination between the disk and companion orbit is damped by viscous dissipation, as
\begin{equation}
t_{\rm a}
=
\frac{(h/r)^2\Omega_{\rm comp}}
{\alpha \omega_{\rm n,global}^{\,2}},
\end{equation}
where $\alpha$ is the viscosity parameter, and $\omega_{\rm n,global}=2\pi/t_{\rm n,global}$ is the global nodal precession angular frequency. For a conservative estimate of the alignment timescale, we use the disk aspect ratio evaluated at $r_{\rm in}=20~{\rm au}$, $(h/r)_{\rm in}=0.119$ \citep{UbeiraGabellini2019}, and assume $\alpha=0.01$ when estimating $t_{\rm a}$ over the parameter space of companion orbital semimajor axis and mass. The resulting map is shown in Figure~\ref{fig:ta_map}. In most of the explored parameter space, $t_{\rm a}$ exceeds 10 Myr. Adopting a lower viscosity would further lengthen the alignment timescale. Since CQ~Tau is of order 10 Myr old \citep{Garufi2024,Wallack2024}, this suggests that a warp induced by an inclined companion could persist over a timescale comparable to or longer than the system age. Therefore, the companion-driven warp scenario is plausible as an origin of the warp in the dust disk of CQ~Tau.

A companion can also produce other structures, such as a cavity \citep{Dong2015b}. CQ~Tau is known to host a central cavity of $\sim 20$ au \citep{UbeiraGabellini2019}, which is qualitatively consistent with the idea that a companion can carve the disk. At the same time, caution is required when linking the inferred companion directly to the full cavity size. Indeed, \citet{Vioque2026} argued that a single companion is unlikely to account for the full cavity size in CQ~Tau. This suggests that additional processes would be required to explain the observed cavity, although a companion can still contribute to shaping the overall disk architecture.

A companion can also produce spiral arms \citep{Dong2015a} and a misaligned inner disk \citep{Zhu2019}. The companion scenario is therefore qualitatively consistent with the multiple structures observed in the CQ~Tau disk, including the NIR spiral arms and the misaligned inner disk \citep{Bohn2022,Hammond2022}. In addition, the shadows cast by such a misaligned inner disk may further induce warps in the outer disk \citep{Zhang2025}. Overall, the companion scenario may provide a natural explanation for the ring-scale warp in CQ~Tau.

%====================================Fig====================================
%====================================Fig====================================

\subsubsection{Flyby}
\label{sec:warporigin_flyby}
A stellar flyby can also induce warps with amplitudes of a few degrees in the outer disk \citep{Kimmig2026}. Such amplitudes are broadly comparable to the moderate warp inferred for the CQ Tau disk. However, such perturbations are not expected to persist indefinitely. In low-viscosity simulations \citep{Kimmig2026}, flyby-induced warps survive on a timescale comparable to the viscous damping timescale $t_{\rm damp} \sim 1/(\alpha \Omega_{\rm out})$ expected from linear bending-wave theory \citep{Lubow2000}, where $\alpha$ is the viscosity parameter and $\Omega_{\rm out}$ is the Keplerian angular frequency at the outer disk radius. Rescaling this damping timescale to the CQ~Tau case at $r_{\rm out}=70~\rm au$ yields values ranging from $\sim10^4$ to $\sim10^6$ yr over the representative range $\alpha=10^{-4}$--$10^{-2}$ \citep{Rosotti2023}. While CQ~Tau does not show clear interaction with the surrounding environment \citep{Garufi2024}, a recent flyby cannot be ruled out as a possible origin of the warp, particularly if the disk viscosity is sufficiently low.

\subsubsection{Late Infall}
\label{sec:warporigin_infall}
Late infall of material with misaligned angular momentum can also tilt disks and change their angular momentum direction \citep[e.g.,][]{Bate2010,Kuffmeier2021}. Indeed, several young systems showing warps have also been reported to exhibit asymmetric infall or streamer-like features \citep{Sai2020,Ginski2021,Tanious2024}. On the other hand, CQ~Tau is a relatively evolved system with an age of order 10 Myr \citep{Garufi2024,Wallack2024} and shows no clear evidence for ongoing large-scale asymmetric infall. However, a contribution from a past infall event cannot be ruled out, as simulations suggest that infall-induced inner--outer disk misalignment can survive for $\sim 0.1~{\rm Myr}$ or longer \citep{Kuffmeier2021}.

\subsubsection{Magnetic Torque}
\label{sec:warporigin_magnetictorque}
Magnetic torque arising from a misalignment between the stellar magnetic field and the disk rotation axis has been proposed as one mechanism for producing warps \citep{Lai1999,Foucart2011}. These studies mainly describe how magnetic warping is induced in the disk inner region and communicated through a continuously connected disk. In contrast, the CQ~Tau system contains a misaligned inner disk \citep{Bohn2022}, a cavity, and the warped main ring. Because magnetic torques are expected to act most effectively in the innermost disk regions, they may be relevant to the origin of the misalignment of the sub-au-scale inner disk in CQ~Tau, but they do not appear to provide a straightforward explanation for the warp in the main ring.

\section{Summary}
\label{sec:summary}

In this study, we combined three archival ALMA Band 6 datasets of CQ~Tau and analyzed the dust disk by applying the super-resolution imaging method PRIISM to the 1.3 mm dust continuum data. The reconstructed image reveals a compact unresolved inner disk and a bright main ring extending from $\sim 20$ to $\sim 70$ au. 
\rev{The PRIISM image provides a higher imaging fidelity than that of the CLEAN images and achieves an effective spatial resolution of $0.052''\times0.042''$, comparable to that of the CLEAN image with a robust parameter of -0.2.}
To examine the underlying disk geometry while reducing the influence of non-centro-symmetric structures, we also reconstructed \rev{PRIISM and CLEAN (robust = $-0.2$) images} using only the real part of the visibilities. The main scientific results are summarized as follows.

\begin{enumerate}
\item
\rev{Ellipse fitting applied to multiple contours of the real-part PRIISM image with the ellipse center fixed at the stellar position reveals radial variations in PA and $i$ at the inner and outer edges of the main ring. At the inner edge ($\sim21$--$26$ au), PA and $i$ span $214^\circ$--$233^\circ$ and $23^\circ$--$39^\circ$, respectively, while they span $231^\circ$--$235^\circ$ and $33^\circ$--$36^\circ$ at the outer edge ($\sim58$--$67$ au).
These results obtained from the PRIISM image reconstructed from the combined multi-epoch data are consistent with those obtained from the CLEAN image with robust = $-0.2$ and the PRIISM image using only the single observing epoch. These results support the robustness of the PA and $i$ variations inside and outside the main dust ring.}

\item
Toy-model tests show that the coplanar ring model cannot reproduce the observed variations in PA and $i$, whereas the fiducial warped ring model provides a better reproduction of the observed radial variations in both quantities. Ellipse fitting including the ellipse center as a free parameter still provides a similar result and the ellipse center is consistent with the stellar position. This suggests that the ring is unlikely to trace an eccentric orbit. These results identify the main dust ring of CQ~Tau as warped.

\item
The high-pass-filtered real-part PRIISM image shows a pair of centro-symmetric but non-axisymmetric residual patterns. Similar curved negative residuals appear in the high-pass-filtered warped ring model, whereas they are absent from the coplanar ring model. This comparison suggests that high-pass filtering may provide a supplementary tool for identifying disk warps.

\item
Together with the previously reported strongly misaligned inner disk and candidate gas warp, the dust warp identified in this work suggests that CQ~Tau has a hierarchical three-dimensional disk structure. The disk geometry changes radially from a strongly misaligned inner disk with a tilt of $\sim40^\circ$ to a more weakly warped main dust ring in which $\beta$ decreases from $\sim4^\circ$ at 20 au to $\sim1^\circ$ at 60 au, together with a candidate gas warp with a tilt of a few degrees at $r = 45$ to 155 au.

\item
As possible origins of the warp, a stellar flyby and late infall cannot be ruled out depending on the parameters that control the survival timescales of the induced warps. On the other hand, an inclined companion provides a plausible explanation for the warp and is also consistent with several other structures in CQ~Tau, including the misaligned inner disk, central cavity, and NIR spiral arms.

\end{enumerate}

Taken together, these results establish CQ~Tau as a clear example of a warped dust disk. Combined with the previously reported misaligned inner disk and candidate gas warp, CQ~Tau provides an important laboratory for investigating the connection between inner-disk misalignment, ring-scale warps, and planet formation.

% \begin{acknowledgments}
%% Also note that the akcnowlodgment environment does not support long amounts of text. If you have a lot of people and institutions to acknowledge, do not use this command. Instead, create a new \section{Acknowledgments}.
\section*{Acknowledgments}

We thank the referee for the constructive comments and suggestions, which helped improve the manuscript.
We are grateful to A. Shoshi for valuable comments. We also thank the ALMA Helpdesk staff for their assistance in providing the calibrated measurement sets. This work is part of the ASIAA Summer Student Program 2024, and S.I. acknowledges support from the Academia Sinica Institute of Astronomy and Astrophysics (ASIAA) in Taiwan.
S.I. was supported by the ALMA Japan Research Grant of NAOJ ALMA Project, NAOJ-ALMA-406.
M.Y. acknowledges support from JSPS KAKENHI grant Nos. JP26K17220 and JP26K00741 and from NAOJ ALMA Scientific Research grant Code 2022-22B. H.-W.Y. acknowledges support from the National Science and Technology Council (NSTC) in Taiwan through grant NSTC 115-2112-M-001-006- and from the Academia Sinica Career Development Award (AS-CDA-111-M03).
S.T. is supported by JSPS KAKENHI grant Nos. JP21H04495 and JP24K00674 and by NAOJ ALMA Scientific Research grant No. 2022-20A. 
Data analysis was in part carried out on the Multi-wavelength Data Analysis System operated by the Astronomy Data Center (ADC), National Astronomical Observatory of Japan.
This paper makes use of the following ALMA data:
ADS/JAO.ALMA\allowbreak\#2013.1.00498.S, ADS/JAO.ALMA\allowbreak\#2016.A.00026.S, and ADS/JAO.ALMA\allowbreak\#2017.1.01404.S.
ALMA is a partnership of ESO (representing its member states), NSF (USA) and NINS (Japan), together with NRC (Canada), NSTC and ASIAA (Taiwan), and KASI (Republic of Korea), in cooperation with the Republic of Chile. The Joint ALMA Observatory is operated by ESO, AUI/NRAO and NAOJ.

% \end{acknowledgments}


% \vspace{5mm}
\facility{ALMA}


\software{
\rev{AJISAI v0.1.0 \citep{Yamaguchi2026AJISAI}}, 
Astropy \citep{Astropy2013,Astropy2018,Astropy2022}, 
CASA \citep{CASA2022}, 
emcee \citep{emcee2013}, 
matplotlib \citep{Matplotlib2007}, 
Plotly \citep{plotly}, 
PRIISM \citep{Nakazato2020,Nakazato2020_SPIE}, 
SciPy \citep{SciPy2020}}
% Shapely \citep{Shapely2025}



%##################################Appendix#################################

\appendix
\restartappendixnumbering

\section{Procedure for Ellipse Fitting}
\label{appendix:ellipsefit}

In this section, we describe the ellipse-fitting procedure applied to the observed emission contours. In the center-fixed case, the ellipse center is fixed at the phase center by setting $(x_c,y_c)=(0,0)$, where $(x_c,y_c)$ denotes the offset of the ellipse center from the phase center. The free parameters are therefore $(x_c,~y_c,~a,~b,~\mathrm{PA})$ for the center-free fit and $(a,~b,~\mathrm{PA})$ for the center-fixed fit, and the fitting is performed using MCMC with \texttt{emcee} \citep{emcee2013}.

For each contour, we first extract a set of coordinates $(x_j,~y_j)$ along the contour from \rev{the corresponding continuum} image using \texttt{matplotlib} \citep{Matplotlib2007}. Because adjacent contour points are sampled more densely than the spatial resolution of the image, they are not treated as independent measurements. Instead, consecutive contour points are grouped along the contour within a cumulative distance \rev{corresponding to the spatial-resolution scale, taken as the major-axis FWHM.} Once this distance is exceeded, a new grouping is triggered. 
\rev{This grouping also reduces the influence of noise-induced local fluctuations in the contours. However, as shown in Figure~\ref{fig:ellipsefit_bfdpj_centerfixed_realonly}, the observed contours already show clear elliptical shapes without significant fluctuations. We therefore adopt the spatial resolution as the grouping scale.}
For each group, the representative data point $(x_i,~y_i)$ is defined as the mean position of the contour points included in that group. To this data point, the positional uncertainties, $(\sigma_{x_i},~\sigma_{y_i})$, are defined as the standard deviations of the grouped coordinates in the $x$ and $y$ directions, respectively. These uncertainties are used as effective positional uncertainties that account for the finite \rev{spatial} resolution and the local spread of contour points within each spatial-resolution-scale segment.

The derived data points $(x_i,~y_i)$ are then converted to polar coordinates $(r_{\mathrm{data},i},~\mathrm{PA}_i)$ with respect to the ellipse center $(x_c,~y_c)$. The observed radial distance $r_{\mathrm{data},i}$ is defined as
\begin{equation}
r_{\mathrm{data},i}
= \sqrt{(x_i - x_c)^2 + (y_i - y_c)^2}.
\end{equation}
For the center-free fit, this coordinate conversion is recalculated at each MCMC sampling step.
On the other hand, the radial distance of the model ellipse at a given position angle $\mathrm{PA}_i$, denoted as $r_{\mathrm{model},i}$, is derived analytically:
\begin{equation}
r_{\mathrm{model},i}
= \frac{ab}{
\sqrt{
b^2 \cos^2(\mathrm{PA}_i - \mathrm{PA})
+ a^2 \sin^2(\mathrm{PA}_i - \mathrm{PA})
}
}.
\end{equation}
The residual for each derived data point is defined as the difference between the observed radius $r_{\mathrm{data},i}$ and the model radius $r_{\mathrm{model},i}$. 
Observational uncertainties in the data point positions, $(\sigma_{x_i},~\sigma_{y_i})$, are propagated to estimate the uncertainty in the radial direction, $\sigma_{r_i}$, based on the standard first-order error propagation formula. The log-likelihood function for the ellipse fitting is defined as
\begin{equation}
\ln \mathcal{L}
= -\frac{1}{2}
\sum_i
\left\{
\left(
\frac{r_{\mathrm{data},i} - r_{\mathrm{model},i}}{\sigma_{r_i}}
\right)^2
+ \ln \left( 2\pi \sigma_{r_i}^2 \right)
\right\}.
\end{equation}
For the center-free fit, $\sigma_{r_i}$ varies with $(x_c,y_c)$ and is therefore updated at each MCMC sampling step. Based on this likelihood function, we estimate the posterior distributions of the free parameters and adopt the median values and the 16th–84th percentile ranges as the representative values and uncertainties. The inclination $i$ is derived from the sampled ellipse parameters, $a$ and $b$, as $i=\arccos(b/a)$.

To assess the goodness of fit, we also compute the reduced chi-square,
\begin{equation}
\chi_{\rm red}^2
=
\frac{1}{N_{\rm data}-N_{\rm par}}
\sum_i
\left(
\frac{r_{\mathrm{data},i} - r_{\mathrm{model},i}}{\sigma_{r_i}}
\right)^2,
\end{equation}
where $N_{\rm data}$ is the number of grouped data points and $N_{\rm par}$ is the number of free parameters. This quantity is evaluated for the median posterior parameter set.

\section{Construction of the Coplanar Model}
\label{appendix:make_basetoy}
\resetapptablenumbers
% ====================================Fig====================================
\begin{figure}[t]
\centering
\includegraphics[width=0.46\textwidth]{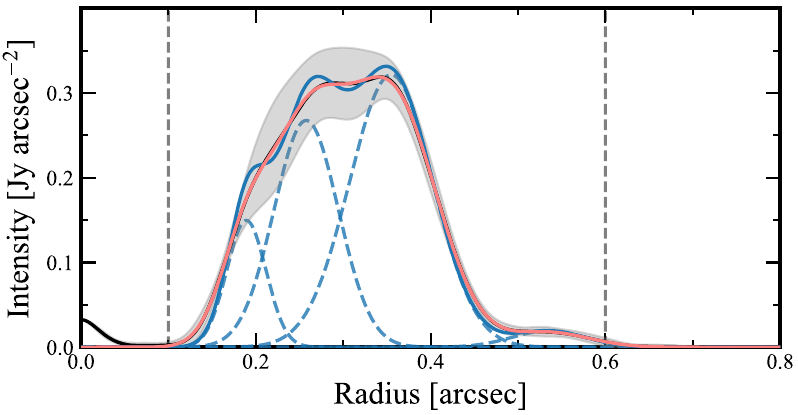}
\caption{
Radial profile adopted to construct the coplanar ring model. The black curve and gray band show the azimuthally averaged radial profile of the deprojected real-part PRIISM image and its standard deviation at each radius, respectively. The red curve shows the best-fit convolved multi-Gaussian profile. The blue solid curve shows the intrinsic multi-Gaussian profile, while the blue dashed curves show its individual Gaussian components. The vertical dashed lines indicate the fitting range.
}
\label{fig:par_profile}
\end{figure}
%====================================Fig====================================

To construct an infinitesimally thin coplanar ring model, we first deprojected the real-part PRIISM image using PA$_0$ and $i_0$ and then computed an azimuthally averaged radial intensity profile (Figure~\ref{fig:par_profile}). The uncertainty at each radius was estimated as the standard deviation over all azimuthal angles. 

Next, we modeled the radial intensity profile as a sum of Gaussian components,
\begin{equation}
I(r) = \sum_{k=1}^{N_{\rm G}}
A_k \exp\left[-\frac{(r-\mu_k)^2}{2\sigma_k^2}\right],
\end{equation}
where $A_k$, $\mu_k$, and $\sigma_k$ are the peak intensity, central radius, and radial width of the $k$th Gaussian component, respectively, and $N_{\rm G}$ is the number of Gaussian components. This multi-Gaussian profile was convolved with a one-dimensional Gaussian whose FWHM was set to the geometric mean of the effective spatial resolution of the PRIISM image, and the convolved profile was fitted to the observed radial profile using the \texttt{curve\_fit} function in \texttt{scipy.optimize} (Figure~\ref{fig:par_profile}). The fitting was performed in the radial range of $0.1''$--$0.6''$ to avoid contamination from the inner compact disk emission. The number of Gaussian components was determined as the minimum number for which the absolute residuals between the observed and convolved model profiles remained below $I_{\rm DT}$ throughout the fitting range. This criterion resulted in $N_{\rm G}=4$, and the corresponding best-fit parameters are summarized in Table~\ref{tab:multigaussian_params}.

Then, the best-fit intrinsic profile was azimuthally rotated to construct a two-dimensional face-on axisymmetric ring model. The face-on model was subsequently projected onto the sky plane using PA$_0$ and $i_0$ while conserving the total flux, as appropriate for the observed optically thin continuum emission.

%-----------------------------Table Start-----------------------------
\begin{table}[t]
\centering
\refstepcounter{table}
{\centering \textbf{Table \thetable}\par}
{\centering Best-fit parameters of the intrinsic multi-Gaussian profile\par}
\label{tab:multigaussian_params}
\vspace{3pt}

\renewcommand{\arraystretch}{1.1}
\begin{tabular*}{\columnwidth}{
    @{\extracolsep{\fill}}
    l
    c
    c
    c
    @{}
}
\hline\hline
 & $A_k$ [$\rm Jy~arcsec^{-2}$]
 & $\mu_k$ [$\rm arcsec$]
 & $\sigma_k$ [$\rm arcsec$] \\
\hline
Component 1 & 0.150 & 0.189 & 0.0225 \\
Component 2 & 0.268 & 0.258 & 0.0369 \\
Component 3 & 0.322 & 0.355 & 0.0478 \\
Component 4 & 0.0194 & 0.535 & 0.0414 \\
\hline
\end{tabular*}

% \vspace{2pt}
% \tablecomments{\parbox{\columnwidth}{
% \textbf{Note.} xxx
% }}
\end{table}
%------------------------Table End ------------------------------

\section{Dependence of the High-pass-filtered Images on the Smoothing Scale}
\label{appendix:highpass}
\resetapptablenumbers

%====================================Fig====================================
\begin{figure*}[t]
\centering
\includegraphics[width=1.0\textwidth]{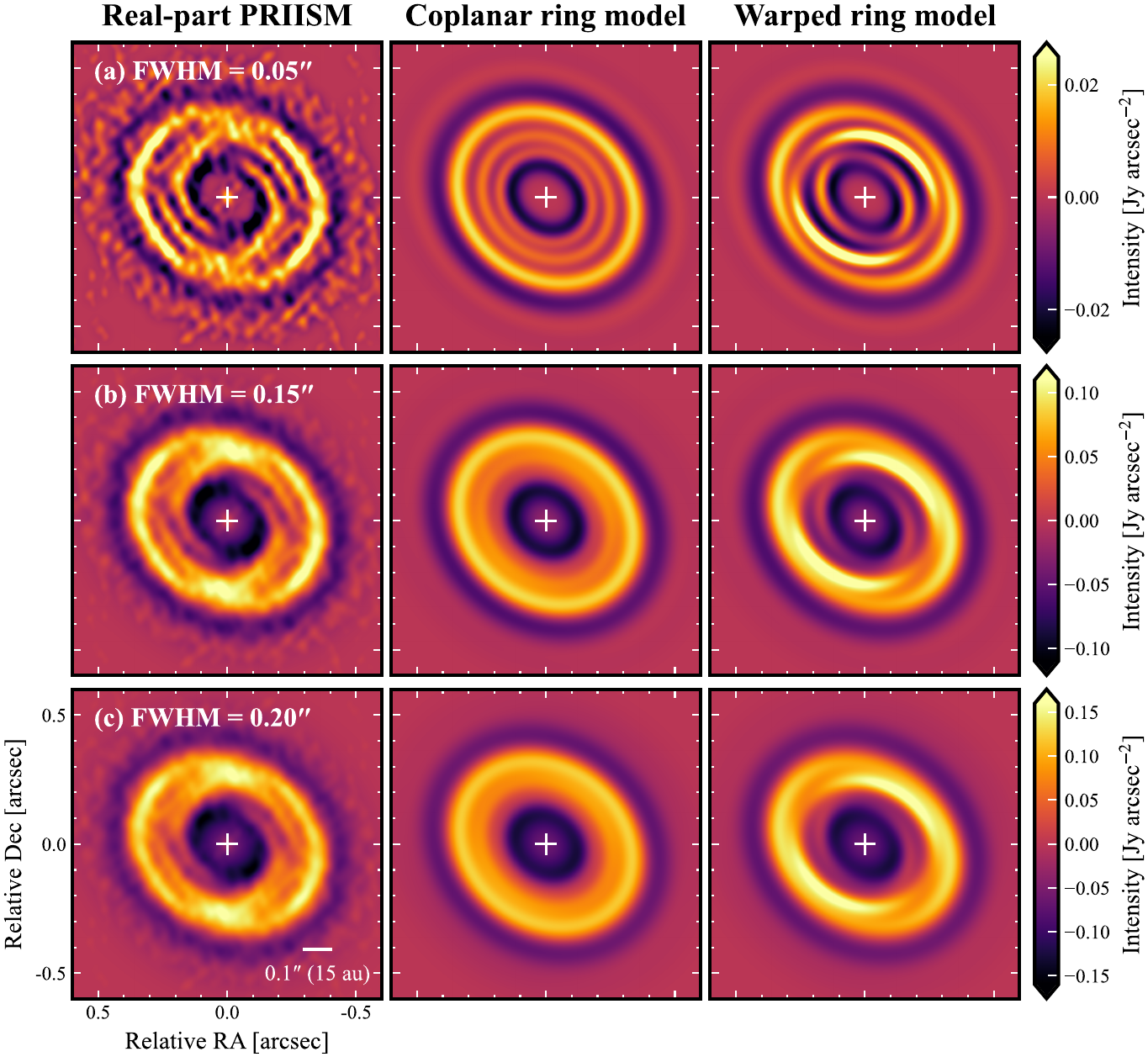}
\caption{
Same as Figure~\ref{fig:gallery_HpFwhm0p1_main}, but for smoothing FWHMs of $0.05''$, $0.15''$, and $0.20''$ from top to bottom. The color scale is common to the three panels within each row but differs between rows.
}
\label{fig:gallery_highpass}
\end{figure*}
%====================================Fig====================================

Figure~\ref{fig:gallery_highpass} compares the high-pass-filtered images obtained with smoothing FWHMs of $0.05''$, $0.15''$, and $0.20''$, while Figure~\ref{fig:gallery_HpFwhm0p1_main} shows the images obtained with the adopted FWHM of $0.1''$. Although the detailed appearance and residual intensity vary with the smoothing FWHM, the qualitative differences among the real-part PRIISM image, coplanar ring model, and warped ring model remain unchanged.

\bibliography{Reference}{}

@software{Yamaguchi2026AJISAI,
  author       = {Masayuki Yamaguchi},
  title        = {AJISAI: Automated Justification-based Imaging and
                   Self-calibration for ALMA Infrastructure
                  },
  month        = sep,
  year         = 2026,
  publisher    = {Zenodo},
  version      = {v0.1.0},
  doi          = {10.5281/zenodo.23053993},
  url          = {https://doi.org/10.5281/zenodo.23053993},
  swhid        = {swh:1:dir:e0301fc06a362485c4ce52b3eeb64852e4b3563f
                   ;origin=https://doi.org/10.5281/zenodo.23053992;vi
                   sit=swh:1:snp:b0f80df48b28c0e8f98dc09f7c25d4d13d26
                   6410;anchor=swh:1:rel:4751679e44e85ece3f6ead57acb0
                   357ec7917879;path=Y-Masayuki-AJISAI-313eaf8
                  },
}

@ARTICLE{Beckwith1990,
       author = {{Beckwith}, Steven V.~W. and {Sargent}, Anneila I. and {Chini}, Rolf S. and {Guesten}, Rolf},
        title = "{A Survey for Circumstellar Disks around Young Stellar Objects}",
      journal = {\aj},
         year = 1990,
        month = mar,
       volume = {99},
        pages = {924},
          doi = {10.1086/115385},
       adsurl = {https://ui.adsabs.harvard.edu/abs/1990AJ.....99..924B}
}

@ARTICLE{Albrecht2022,
       author = {{Albrecht}, Simon H. and {Dawson}, Rebekah I. and {Winn}, Joshua N.},
        title = "{Stellar Obliquities in Exoplanetary Systems}",
      journal = {\pasp},
         year = 2022,
        month = aug,
       volume = {134},
       number = {1038},
          eid = {082001},
        pages = {082001},
          doi = {10.1088/1538-3873/ac6c09},
archivePrefix = {arXiv},
       eprint = {2203.05460},
 primaryClass = {astro-ph.EP},
       adsurl = {https://ui.adsabs.harvard.edu/abs/2022PASP..134h2001A}
}

@article{Rosotti2023,
title = {Empirical constraints on turbulence in proto-planetary discs},
journal = {New Astronomy Reviews},
volume = {96},
pages = {101674},
year = {2023},
issn = {1387-6473},
doi = {https://doi.org/10.1016/j.newar.2023.101674},
url = {https://www.sciencedirect.com/science/article/pii/S1387647323000027},
author = {Giovanni P. Rosotti}
}

@INPROCEEDINGS{Nakazato2020_SPIE,
       author = {{Nakazato}, Takeshi and {Ikeda}, Shiro and {Kosugi}, George and {Honma}, Mareki},
        title = "{PRIISM: Synthesis imaging tool based on the sparse modeling for radio astronomy}",
    booktitle = {Millimeter, Submillimeter, and Far-Infrared Detectors and Instrumentation for Astronomy X},
         year = 2020,
       editor = {{Zmuidzinas}, Jonas and {Gao}, Jian-Rong},
       series = {Society of Photo-Optical Instrumentation Engineers (SPIE) Conference Series},
       volume = {11453},
        month = dec,
          eid = {114532V},
        pages = {114532V},
          doi = {10.1117/12.2560904},
       adsurl = {https://ui.adsabs.harvard.edu/abs/2020SPIE11453E..2VN}
}

@ARTICLE{Casassus2018,
       author = {{Casassus}, Simon and {Avenhaus}, Henning and {P{\'e}rez}, Sebasti{\'a}n and {Navarro}, V{\'\i}ctor and {C{\'a}rcamo}, Miguel and {Marino}, Sebasti{\'a}n and {Cieza}, Lucas and {Quanz}, Sascha P. and {Alarc{\'o}n}, Felipe and {Zurlo}, Alice and {Osses}, Axel and {Rannou}, Fernando R. and {Rom{\'a}n}, Pablo E. and {Barraza}, Marcelo},
        title = "{An inner warp in the DoAr 44 T Tauri transition disc}",
      journal = {\mnras},
         year = 2018,
        month = jul,
       volume = {477},
       number = {4},
        pages = {5104-5114},
          doi = {10.1093/mnras/sty894},
archivePrefix = {arXiv},
       eprint = {1804.02360},
 primaryClass = {astro-ph.SR},
       adsurl = {https://ui.adsabs.harvard.edu/abs/2018MNRAS.477.5104C}
}

@ARTICLE{Kuffmeier2021,
       author = {{Kuffmeier}, M. and {Dullemond}, C.~P. and {Reissl}, S. and {Goicovic}, F.~G.},
        title = "{Misaligned disks induced by infall}",
      journal = {\aap},
         year = 2021,
        month = dec,
       volume = {656},
          eid = {A161},
        pages = {A161},
          doi = {10.1051/0004-6361/202039614},
archivePrefix = {arXiv},
       eprint = {2110.04309},
 primaryClass = {astro-ph.SR},
       adsurl = {https://ui.adsabs.harvard.edu/abs/2021A&A...656A.161K}
}

@ARTICLE{GaiaCollaboration2023,
       author = {{Gaia Collaboration} and {Vallenari}, A. and {Brown}, A.~G.~A. and {Prusti}, T. and {de Bruijne}, J.~H.~J. and {Arenou}, F. and {Babusiaux}, C. and {Biermann}, M. and {Creevey}, O.~L. and {Ducourant}, C. and {Evans}, D.~W. and {Eyer}, L. and {Guerra}, R. and {Hutton}, A. and {Jordi}, C. and {Klioner}, S.~A. and {Lammers}, U.~L. and {Lindegren}, L. and {Luri}, X. and {Mignard}, F. and {Panem}, C. and {Pourbaix}, D. and {Randich}, S. and {Sartoretti}, P. and {Soubiran}, C. and {Tanga}, P. and {Walton}, N.~A. and {Bailer-Jones}, C.~A.~L. and {Bastian}, U. and {Drimmel}, R. and {Jansen}, F. and {Katz}, D. and {Lattanzi}, M.~G. and {van Leeuwen}, F. and {Bakker}, J. and {Cacciari}, C. and {Casta{\~n}eda}, J. and {De Angeli}, F. and {Fabricius}, C. and {Fouesneau}, M. and {Fr{\'e}mat}, Y. and {Galluccio}, L. and {Guerrier}, A. and {Heiter}, U. and {Masana}, E. and {Messineo}, R. and {Mowlavi}, N. and {Nicolas}, C. and {Nienartowicz}, K. and {Pailler}, F. and {Panuzzo}, P. and {Riclet}, F. and {Roux}, W. and {Seabroke}, G.~M. and {Sordo}, R. and {Th{\'e}venin}, F. and {Gracia-Abril}, G. and {Portell}, J. and {Teyssier}, D. and {Altmann}, M. and {Andrae}, R. and {Audard}, M. and {Bellas-Velidis}, I. and {Benson}, K. and {Berthier}, J. and {Blomme}, R. and {Burgess}, P.~W. and {Busonero}, D. and {Busso}, G. and {C{\'a}novas}, H. and {Carry}, B. and {Cellino}, A. and {Cheek}, N. and {Clementini}, G. and {Damerdji}, Y. and {Davidson}, M. and {de Teodoro}, P. and {Nu{\~n}ez Campos}, M. and {Delchambre}, L. and {Dell'Oro}, A. and {Esquej}, P. and {Fern{\'a}ndez-Hern{\'a}ndez}, J. and {Fraile}, E. and {Garabato}, D. and {Garc{\'\i}a-Lario}, P. and {Gosset}, E. and {Haigron}, R. and {Halbwachs}, J.-L. and {Hambly}, N.~C. and {Harrison}, D.~L. and {Hern{\'a}ndez}, J. and {Hestroffer}, D. and {Hodgkin}, S.~T. and {Holl}, B. and {Jan{\ss}en}, K. and {Jevardat de Fombelle}, G. and {Jordan}, S. and {Krone-Martins}, A. and {Lanzafame}, A.~C. and {L{\"o}ffler}, W. and {Marchal}, O. and {Marrese}, P.~M. and {Moitinho}, A. and {Muinonen}, K. and {Osborne}, P. and {Pancino}, E. and {Pauwels}, T. and {Recio-Blanco}, A. and {Reyl{\'e}}, C. and {Riello}, M. and {Rimoldini}, L. and {Roegiers}, T. and {Rybizki}, J. and {Sarro}, L.~M. and {Siopis}, C. and {Smith}, M. and {Sozzetti}, A. and {Utrilla}, E. and {van Leeuwen}, M. and {Abbas}, U. and {{\'A}brah{\'a}m}, P. and {Abreu Aramburu}, A. and {Aerts}, C. and {Aguado}, J.~J. and {Ajaj}, M. and {Aldea-Montero}, F. and {Altavilla}, G. and {{\'A}lvarez}, M.~A. and {Alves}, J. and {Anders}, F. and {Anderson}, R.~I. and {Anglada Varela}, E. and {Antoja}, T. and {Baines}, D. and {Baker}, S.~G. and {Balaguer-N{\'u}{\~n}ez}, L. and {Balbinot}, E. and {Balog}, Z. and {Barache}, C. and {Barbato}, D. and {Barros}, M. and {Barstow}, M.~A. and {Bartolom{\'e}}, S. and {Bassilana}, J.-L. and {Bauchet}, N. and {Becciani}, U. and {Bellazzini}, M. and {Berihuete}, A. and {Bernet}, M. and {Bertone}, S. and {Bianchi}, L. and {Binnenfeld}, A. and {Blanco-Cuaresma}, S. and {Blazere}, A. and {Boch}, T. and {Bombrun}, A. and {Bossini}, D. and {Bouquillon}, S. and {Bragaglia}, A. and {Bramante}, L. and {Breedt}, E. and {Bressan}, A. and {Brouillet}, N. and {Brugaletta}, E. and {Bucciarelli}, B. and {Burlacu}, A. and {Butkevich}, A.~G. and {Buzzi}, R. and {Caffau}, E. and {Cancelliere}, R. and {Cantat-Gaudin}, T. and {Carballo}, R. and {Carlucci}, T. and {Carnerero}, M.~I. and {Carrasco}, J.~M. and {Casamiquela}, L. and {Castellani}, M. and {Castro-Ginard}, A. and {Chaoul}, L. and {Charlot}, P. and {Chemin}, L. and {Chiaramida}, V. and {Chiavassa}, A. and {Chornay}, N. and {Comoretto}, G. and {Contursi}, G. and {Cooper}, W.~J. and {Cornez}, T. and {Cowell}, S. and {Crifo}, F. and {Cropper}, M. and {Crosta}, M. and {Crowley}, C. and {Dafonte}, C. and {Dapergolas}, A. and {David}, M. and {David}, P. and {de Laverny}, P. and {De Luise}, F. and {De March}, R.},
        title = "{Gaia Data Release 3. Summary of the content and survey properties}",
      journal = {\aap},
         year = 2023,
        month = jun,
       volume = {674},
          eid = {A1},
        pages = {A1},
          doi = {10.1051/0004-6361/202243940},
archivePrefix = {arXiv},
       eprint = {2208.00211},
 primaryClass = {astro-ph.GA},
       adsurl = {https://ui.adsabs.harvard.edu/abs/2023A&A...674A...1G}
}

@ARTICLE{Yamaguchi2026,
       author = {{Yamaguchi}, Masayuki and {Machida}, Masahiro N. and {Tominaga}, Ryosuke T. and {Sai}, Jinshi and {Muto}, Takayuki and {Takami}, Michihiro and {Liu}, Hauyu Baobab and {Shoshi}, Ayumu and {Tsukagoshi}, Takashi and {Ishibashi}, Shu},
        title = "{A Hybrid Origin for the Multiple Ring-gap Structures in the Large Protoplanetary Disk V1094 Sco: A Low-mass Planet and Secular Gravitational Instability}",
      journal = {\apj},
         year = 2026,
        month = aug,
       volume = {1006},
       number = {2},
          eid = {232},
        pages = {232},
          doi = {10.3847/1538-4357/ae819b},
archivePrefix = {arXiv},
       eprint = {2605.11486},
 primaryClass = {astro-ph.EP},
       adsurl = {https://ui.adsabs.harvard.edu/abs/2026ApJ..1006..232Y}
}

@software{plotly,
  author = {{Plotly Technologies Inc.}},
  title = {Plotly},
  year = {2015},
  url = {https://plotly.com}
}

@ARTICLE{SciPy2020,
  author  = {Virtanen, Pauli and Gommers, Ralf and Oliphant, Travis E. and
            Haberland, Matt and Reddy, Tyler and Cournapeau, David and
            Burovski, Evgeni and Peterson, Pearu and Weckesser, Warren and
            Bright, Jonathan and {van der Walt}, St{\'e}fan J. and
            Brett, Matthew and Wilson, Joshua and Millman, K. Jarrod and
            Mayorov, Nikolay and Nelson, Andrew R. J. and Jones, Eric and
            Kern, Robert and Larson, Eric and Carey, C J and
            Polat, {\.I}lhan and Feng, Yu and Moore, Eric W. and
            {VanderPlas}, Jake and Laxalde, Denis and Perktold, Josef and
            Cimrman, Robert and Henriksen, Ian and Quintero, E. A. and
            Harris, Charles R. and Archibald, Anne M. and
            Ribeiro, Ant{\^o}nio H. and Pedregosa, Fabian and
            {van Mulbregt}, Paul and {SciPy 1.0 Contributors}},
  title   = {{{SciPy} 1.0: Fundamental Algorithms for Scientific
            Computing in Python}},
  journal = {Nature Methods},
  year    = {2020},
  volume  = {17},
  pages   = {261--272},
  adsurl  = {https://rdcu.be/b08Wh},
  doi     = {10.1038/s41592-019-0686-2},
}

@ARTICLE{Matplotlib2007,
  author={Hunter, John D.},
  journal={Computing in Science \& Engineering}, 
  title={Matplotlib: A 2D Graphics Environment}, 
  year={2007},
  volume={9},
  number={3},
  pages={90-95},
  doi={10.1109/MCSE.2007.55}}

@ARTICLE{Astropy2013,
       author = {{Astropy Collaboration} and {Robitaille}, Thomas P. and {Tollerud}, Erik J. and {Greenfield}, Perry and {Droettboom}, Michael and {Bray}, Erik and {Aldcroft}, Tom and {Davis}, Matt and {Ginsburg}, Adam and {Price-Whelan}, Adrian M. and {Kerzendorf}, Wolfgang E. and {Conley}, Alexander and {Crighton}, Neil and {Barbary}, Kyle and {Muna}, Demitri and {Ferguson}, Henry and {Grollier}, Fr{\'e}d{\'e}ric and {Parikh}, Madhura M. and {Nair}, Prasanth H. and {Unther}, Hans M. and {Deil}, Christoph and {Woillez}, Julien and {Conseil}, Simon and {Kramer}, Roban and {Turner}, James E.~H. and {Singer}, Leo and {Fox}, Ryan and {Weaver}, Benjamin A. and {Zabalza}, Victor and {Edwards}, Zachary I. and {Azalee Bostroem}, K. and {Burke}, D.~J. and {Casey}, Andrew R. and {Crawford}, Steven M. and {Dencheva}, Nadia and {Ely}, Justin and {Jenness}, Tim and {Labrie}, Kathleen and {Lim}, Pey Lian and {Pierfederici}, Francesco and {Pontzen}, Andrew and {Ptak}, Andy and {Refsdal}, Brian and {Servillat}, Mathieu and {Streicher}, Ole},
        title = "{Astropy: A community Python package for astronomy}",
      journal = {\aap},
         year = 2013,
        month = oct,
       volume = {558},
          eid = {A33},
        pages = {A33},
          doi = {10.1051/0004-6361/201322068},
archivePrefix = {arXiv},
       eprint = {1307.6212},
 primaryClass = {astro-ph.IM},
       adsurl = {https://ui.adsabs.harvard.edu/abs/2013A&A...558A..33A}
}

@ARTICLE{Astropy2018,
       author = {{Astropy Collaboration} and {Price-Whelan}, A.~M. and {Sip{\H{o}}cz}, B.~M. and {G{\"u}nther}, H.~M. and {Lim}, P.~L. and {Crawford}, S.~M. and {Conseil}, S. and {Shupe}, D.~L. and {Craig}, M.~W. and {Dencheva}, N. and {Ginsburg}, A. and {VanderPlas}, J.~T. and {Bradley}, L.~D. and {P{\'e}rez-Su{\'a}rez}, D. and {de Val-Borro}, M. and {Aldcroft}, T.~L. and {Cruz}, K.~L. and {Robitaille}, T.~P. and {Tollerud}, E.~J. and {Ardelean}, C. and {Babej}, T. and {Bach}, Y.~P. and {Bachetti}, M. and {Bakanov}, A.~V. and {Bamford}, S.~P. and {Barentsen}, G. and {Barmby}, P. and {Baumbach}, A. and {Berry}, K.~L. and {Biscani}, F. and {Boquien}, M. and {Bostroem}, K.~A. and {Bouma}, L.~G. and {Brammer}, G.~B. and {Bray}, E.~M. and {Breytenbach}, H. and {Buddelmeijer}, H. and {Burke}, D.~J. and {Calderone}, G. and {Cano Rodr{\'\i}guez}, J.~L. and {Cara}, M. and {Cardoso}, J.~V.~M. and {Cheedella}, S. and {Copin}, Y. and {Corrales}, L. and {Crichton}, D. and {D'Avella}, D. and {Deil}, C. and {Depagne}, {\'E}. and {Dietrich}, J.~P. and {Donath}, A. and {Droettboom}, M. and {Earl}, N. and {Erben}, T. and {Fabbro}, S. and {Ferreira}, L.~A. and {Finethy}, T. and {Fox}, R.~T. and {Garrison}, L.~H. and {Gibbons}, S.~L.~J. and {Goldstein}, D.~A. and {Gommers}, R. and {Greco}, J.~P. and {Greenfield}, P. and {Groener}, A.~M. and {Grollier}, F. and {Hagen}, A. and {Hirst}, P. and {Homeier}, D. and {Horton}, A.~J. and {Hosseinzadeh}, G. and {Hu}, L. and {Hunkeler}, J.~S. and {Ivezi{\'c}}, {\v{Z}}. and {Jain}, A. and {Jenness}, T. and {Kanarek}, G. and {Kendrew}, S. and {Kern}, N.~S. and {Kerzendorf}, W.~E. and {Khvalko}, A. and {King}, J. and {Kirkby}, D. and {Kulkarni}, A.~M. and {Kumar}, A. and {Lee}, A. and {Lenz}, D. and {Littlefair}, S.~P. and {Ma}, Z. and {Macleod}, D.~M. and {Mastropietro}, M. and {McCully}, C. and {Montagnac}, S. and {Morris}, B.~M. and {Mueller}, M. and {Mumford}, S.~J. and {Muna}, D. and {Murphy}, N.~A. and {Nelson}, S. and {Nguyen}, G.~H. and {Ninan}, J.~P. and {N{\"o}the}, M. and {Ogaz}, S. and {Oh}, S. and {Parejko}, J.~K. and {Parley}, N. and {Pascual}, S. and {Patil}, R. and {Patil}, A.~A. and {Plunkett}, A.~L. and {Prochaska}, J.~X. and {Rastogi}, T. and {Reddy Janga}, V. and {Sabater}, J. and {Sakurikar}, P. and {Seifert}, M. and {Sherbert}, L.~E. and {Sherwood-Taylor}, H. and {Shih}, A.~Y. and {Sick}, J. and {Silbiger}, M.~T. and {Singanamalla}, S. and {Singer}, L.~P. and {Sladen}, P.~H. and {Sooley}, K.~A. and {Sornarajah}, S. and {Streicher}, O. and {Teuben}, P. and {Thomas}, S.~W. and {Tremblay}, G.~R. and {Turner}, J.~E.~H. and {Terr{\'o}n}, V. and {van Kerkwijk}, M.~H. and {de la Vega}, A. and {Watkins}, L.~L. and {Weaver}, B.~A. and {Whitmore}, J.~B. and {Woillez}, J. and {Zabalza}, V. and {Astropy Contributors}},
        title = "{The Astropy Project: Building an Open-science Project and Status of the v2.0 Core Package}",
      journal = {\aj},
         year = 2018,
        month = sep,
       volume = {156},
       number = {3},
          eid = {123},
        pages = {123},
          doi = {10.3847/1538-3881/aabc4f},
archivePrefix = {arXiv},
       eprint = {1801.02634},
 primaryClass = {astro-ph.IM},
       adsurl = {https://ui.adsabs.harvard.edu/abs/2018AJ....156..123A}
}

@ARTICLE{Astropy2022,
       author = {{Astropy Collaboration} and {Price-Whelan}, Adrian M. and {Lim}, Pey Lian and {Earl}, Nicholas and {Starkman}, Nathaniel and {Bradley}, Larry and {Shupe}, David L. and {Patil}, Aarya A. and {Corrales}, Lia and {Brasseur}, C.~E. and {N{\"o}the}, Maximilian and {Donath}, Axel and {Tollerud}, Erik and {Morris}, Brett M. and {Ginsburg}, Adam and {Vaher}, Eero and {Weaver}, Benjamin A. and {Tocknell}, James and {Jamieson}, William and {van Kerkwijk}, Marten H. and {Robitaille}, Thomas P. and {Merry}, Bruce and {Bachetti}, Matteo and {G{\"u}nther}, H. Moritz and {Aldcroft}, Thomas L. and {Alvarado-Montes}, Jaime A. and {Archibald}, Anne M. and {B{\'o}di}, Attila and {Bapat}, Shreyas and {Barentsen}, Geert and {Baz{\'a}n}, Juanjo and {Biswas}, Manish and {Boquien}, M{\'e}d{\'e}ric and {Burke}, D.~J. and {Cara}, Daria and {Cara}, Mihai and {Conroy}, Kyle E. and {Conseil}, Simon and {Craig}, Matthew W. and {Cross}, Robert M. and {Cruz}, Kelle L. and {D'Eugenio}, Francesco and {Dencheva}, Nadia and {Devillepoix}, Hadrien A.~R. and {Dietrich}, J{\"o}rg P. and {Eigenbrot}, Arthur Davis and {Erben}, Thomas and {Ferreira}, Leonardo and {Foreman-Mackey}, Daniel and {Fox}, Ryan and {Freij}, Nabil and {Garg}, Suyog and {Geda}, Robel and {Glattly}, Lauren and {Gondhalekar}, Yash and {Gordon}, Karl D. and {Grant}, David and {Greenfield}, Perry and {Groener}, Austen M. and {Guest}, Steve and {Gurovich}, Sebastian and {Handberg}, Rasmus and {Hart}, Akeem and {Hatfield-Dodds}, Zac and {Homeier}, Derek and {Hosseinzadeh}, Griffin and {Jenness}, Tim and {Jones}, Craig K. and {Joseph}, Prajwel and {Kalmbach}, J. Bryce and {Karamehmetoglu}, Emir and {Ka{\l}uszy{\'n}ski}, Miko{\l}aj and {Kelley}, Michael S.~P. and {Kern}, Nicholas and {Kerzendorf}, Wolfgang E. and {Koch}, Eric W. and {Kulumani}, Shankar and {Lee}, Antony and {Ly}, Chun and {Ma}, Zhiyuan and {MacBride}, Conor and {Maljaars}, Jakob M. and {Muna}, Demitri and {Murphy}, N.~A. and {Norman}, Henrik and {O'Steen}, Richard and {Oman}, Kyle A. and {Pacifici}, Camilla and {Pascual}, Sergio and {Pascual-Granado}, J. and {Patil}, Rohit R. and {Perren}, Gabriel I. and {Pickering}, Timothy E. and {Rastogi}, Tanuj and {Roulston}, Benjamin R. and {Ryan}, Daniel F. and {Rykoff}, Eli S. and {Sabater}, Jose and {Sakurikar}, Parikshit and {Salgado}, Jes{\'u}s and {Sanghi}, Aniket and {Saunders}, Nicholas and {Savchenko}, Volodymyr and {Schwardt}, Ludwig and {Seifert-Eckert}, Michael and {Shih}, Albert Y. and {Jain}, Anany Shrey and {Shukla}, Gyanendra and {Sick}, Jonathan and {Simpson}, Chris and {Singanamalla}, Sudheesh and {Singer}, Leo P. and {Singhal}, Jaladh and {Sinha}, Manodeep and {Sip{\H{o}}cz}, Brigitta M. and {Spitler}, Lee R. and {Stansby}, David and {Streicher}, Ole and {{\v{S}}umak}, Jani and {Swinbank}, John D. and {Taranu}, Dan S. and {Tewary}, Nikita and {Tremblay}, Grant R. and {de Val-Borro}, Miguel and {Van Kooten}, Samuel J. and {Vasovi{\'c}}, Zlatan and {Verma}, Shresth and {de Miranda Cardoso}, Jos{\'e} Vin{\'\i}cius and {Williams}, Peter K.~G. and {Wilson}, Tom J. and {Winkel}, Benjamin and {Wood-Vasey}, W.~M. and {Xue}, Rui and {Yoachim}, Peter and {Zhang}, Chen and {Zonca}, Andrea and {Astropy Project Contributors}},
        title = "{The Astropy Project: Sustaining and Growing a Community-oriented Open-source Project and the Latest Major Release (v5.0) of the Core Package}",
      journal = {\apj},
         year = 2022,
        month = aug,
       volume = {935},
       number = {2},
          eid = {167},
        pages = {167},
          doi = {10.3847/1538-4357/ac7c74},
archivePrefix = {arXiv},
       eprint = {2206.14220},
 primaryClass = {astro-ph.IM},
       adsurl = {https://ui.adsabs.harvard.edu/abs/2022ApJ...935..167A}
}

@ARTICLE{emcee2013,
       author = {{Foreman-Mackey}, Daniel and {Hogg}, David W. and {Lang}, Dustin and {Goodman}, Jonathan},
        title = "{emcee: The MCMC Hammer}",
      journal = {\pasp},
         year = 2013,
        month = mar,
       volume = {125},
       number = {925},
        pages = {306},
          doi = {10.1086/670067},
archivePrefix = {arXiv},
       eprint = {1202.3665},
 primaryClass = {astro-ph.IM},
       adsurl = {https://ui.adsabs.harvard.edu/abs/2013PASP..125..306F}
}

@ARTICLE{CASA2022,
       author = {{CASA Team} and {Bean}, Ben and {Bhatnagar}, Sanjay and {Castro}, Sandra and {Donovan Meyer}, Jennifer and {Emonts}, Bjorn and {Garcia}, Enrique and {Garwood}, Robert and {Golap}, Kumar and {Gonzalez Villalba}, Justo and {Harris}, Pamela and {Hayashi}, Yohei and {Hoskins}, Josh and {Hsieh}, Mingyu and {Jagannathan}, Preshanth and {Kawasaki}, Wataru and {Keimpema}, Aard and {Kettenis}, Mark and {Lopez}, Jorge and {Marvil}, Joshua and {Masters}, Joseph and {McNichols}, Andrew and {Mehringer}, David and {Miel}, Renaud and {Moellenbrock}, George and {Montesino}, Federico and {Nakazato}, Takeshi and {Ott}, Juergen and {Petry}, Dirk and {Pokorny}, Martin and {Raba}, Ryan and {Rau}, Urvashi and {Schiebel}, Darrell and {Schweighart}, Neal and {Sekhar}, Srikrishna and {Shimada}, Kazuhiko and {Small}, Des and {Steeb}, Jan-Willem and {Sugimoto}, Kanako and {Suoranta}, Ville and {Tsutsumi}, Takahiro and {van Bemmel}, Ilse M. and {Verkouter}, Marjolein and {Wells}, Akeem and {Xiong}, Wei and {Szomoru}, Arpad and {Griffith}, Morgan and {Glendenning}, Brian and {Kern}, Jeff},
        title = "{CASA, the Common Astronomy Software Applications for Radio Astronomy}",
      journal = {\pasp},
         year = 2022,
        month = nov,
       volume = {134},
       number = {1041},
          eid = {114501},
        pages = {114501},
          doi = {10.1088/1538-3873/ac9642},
archivePrefix = {arXiv},
       eprint = {2210.02276},
 primaryClass = {astro-ph.IM},
       adsurl = {https://ui.adsabs.harvard.edu/abs/2022PASP..134k4501C}
}

@ARTICLE{Stolker2016,
       author = {{Stolker}, T. and {Dominik}, C. and {Avenhaus}, H. and {Min}, M. and {de Boer}, J. and {Ginski}, C. and {Schmid}, H.~M. and {Juhasz}, A. and {Bazzon}, A. and {Waters}, L.~B.~F.~M. and {Garufi}, A. and {Augereau}, J.-C. and {Benisty}, M. and {Boccaletti}, A. and {Henning}, Th. and {Langlois}, M. and {Maire}, A.-L. and {M{\'e}nard}, F. and {Meyer}, M.~R. and {Pinte}, C. and {Quanz}, S.~P. and {Thalmann}, C. and {Beuzit}, J.-L. and {Carbillet}, M. and {Costille}, A. and {Dohlen}, K. and {Feldt}, M. and {Gisler}, D. and {Mouillet}, D. and {Pavlov}, A. and {Perret}, D. and {Petit}, C. and {Pragt}, J. and {Rochat}, S. and {Roelfsema}, R. and {Salasnich}, B. and {Soenke}, C. and {Wildi}, F.},
        title = "{Shadows cast on the transition disk of HD 135344B. Multiwavelength VLT/SPHERE polarimetric differential imaging}",
      journal = {\aap},
         year = 2016,
        month = nov,
       volume = {595},
          eid = {A113},
        pages = {A113},
          doi = {10.1051/0004-6361/201528039},
archivePrefix = {arXiv},
       eprint = {1603.00481},
 primaryClass = {astro-ph.EP},
       adsurl = {https://ui.adsabs.harvard.edu/abs/2016A&A...595A.113S}
}

@ARTICLE{Winn2015,
       author = {{Winn}, Joshua N. and {Fabrycky}, Daniel C.},
        title = "{The Occurrence and Architecture of Exoplanetary Systems}",
      journal = {\araa},
         year = 2015,
        month = aug,
       volume = {53},
        pages = {409-447},
          doi = {10.1146/annurev-astro-082214-122246},
archivePrefix = {arXiv},
       eprint = {1410.4199},
 primaryClass = {astro-ph.EP},
       adsurl = {https://ui.adsabs.harvard.edu/abs/2015ARA&A..53..409W}
}

@ARTICLE{Orihara2025,
       author = {{Orihara}, Ryuta and {Momose}, Munetake},
        title = "{Shadow-based Framework for Estimating Transition Disk Geometries}",
      journal = {\apj},
         year = 2025,
        month = jun,
       volume = {986},
       number = {2},
          eid = {215},
        pages = {215},
          doi = {10.3847/1538-4357/add890},
archivePrefix = {arXiv},
       eprint = {2505.06044},
 primaryClass = {astro-ph.EP},
       adsurl = {https://ui.adsabs.harvard.edu/abs/2025ApJ...986..215O}
}

@ARTICLE{Vioque2026,
       author = {{Vioque}, Miguel and {Booth}, Richard A. and {Ragusa}, Enrico and {Ribas}, {\'A}lvaro and {Kurtovic}, Nicol{\'a}s T. and {Rosotti}, Giovanni P. and {Penoyre}, Zephyr and {Facchini}, Stefano and {Garufi}, Antonio and {Manara}, Carlo F. and {Hu{\'e}lamo}, Nuria and {Winter}, Andrew and {P{\'e}rez}, Sebasti{\'a}n and {Benisty}, Myriam and {Mendigut{\'\i}a}, Ignacio and {Cuello}, Nicol{\'a}s and {Penzlin}, Anna B.~T. and {Castro-Ginard}, Alfred and {Teague}, Richard},
        title = "{Astrometric view of companions in the inner dust cavities of protoplanetary discs}",
      journal = {\aap},
         year = 2026,
        month = jan,
       volume = {705},
          eid = {A238},
        pages = {A238},
          doi = {10.1051/0004-6361/202557086},
archivePrefix = {arXiv},
       eprint = {2512.00157},
 primaryClass = {astro-ph.EP},
       adsurl = {https://ui.adsabs.harvard.edu/abs/2026A&A...705A.238V}
}

@ARTICLE{Marino2015,
       author = {{Marino}, S. and {Perez}, S. and {Casassus}, S.},
        title = "{Shadows Cast by a Warp in the HD 142527 Protoplanetary Disk}",
      journal = {\apjl},
         year = 2015,
        month = jan,
       volume = {798},
       number = {2},
          eid = {L44},
        pages = {L44},
          doi = {10.1088/2041-8205/798/2/L44},
archivePrefix = {arXiv},
       eprint = {1412.4632},
 primaryClass = {astro-ph.EP},
       adsurl = {https://ui.adsabs.harvard.edu/abs/2015ApJ...798L..44M}
}

@ARTICLE{Izquierdo2025,
       author = {{Izquierdo}, Andr{\'e}s F. and {Stadler}, Jochen and {Galloway-Sprietsma}, Maria and {Benisty}, Myriam and {Pinte}, Christophe and {Bae}, Jaehan and {Teague}, Richard and {Facchini}, Stefano and {W{\"o}lfer}, Lisa and {Longarini}, Cristiano and {Curone}, Pietro and {Andrews}, Sean M. and {Barraza-Alfaro}, Marcelo and {Cataldi}, Gianni and {Cuello}, Nicol{\'a}s and {Czekala}, Ian and {Fasano}, Daniele and {Flock}, Mario and {Fukagawa}, Misato and {Garg}, Himanshi and {Hall}, Cassandra and {Hammond}, Iain and {Hilder}, Thomas and {Huang}, Jane and {Ilee}, John D. and {Isella}, Andrea and {Kanagawa}, Kazuhiro and {Lesur}, Geoffroy and {Lodato}, Giuseppe and {Loomis}, Ryan A. and {Orihara}, Ryuta and {Price}, Daniel J. and {Rosotti}, Giovanni and {Testi}, Leonardo and {Yen}, Hsi-Wei and {Wafflard-Fernandez}, Gaylor and {Wilner}, David J. and {Winter}, Andrew J. and {Yoshida}, Tomohiro C. and {Zawadzki}, Brianna},
        title = "{exoALMA. III. Line-intensity Modeling and System Property Extraction from Protoplanetary Disks}",
      journal = {\apjl},
         year = 2025,
        month = may,
       volume = {984},
       number = {1},
          eid = {L8},
        pages = {L8},
          doi = {10.3847/2041-8213/adc439},
archivePrefix = {arXiv},
       eprint = {2504.19986},
 primaryClass = {astro-ph.EP},
       adsurl = {https://ui.adsabs.harvard.edu/abs/2025ApJ...984L...8I}
}

@ARTICLE{Blakely2026,
       author = {{Blakely}, Dori and {Thompson}, William and {Johnstone}, Doug and {Speedie}, Jessica and {Xuan}, Jerry W. and {Blouin}, Simon and {Zhang}, Jingwen and {Ruffio}, Jean-Baptiste and {Nielsen}, Eric and {Bowler}, Brendan P. and {Franson}, Kyle and {Roberson}, William and {Cloutier}, Ryan and {Fogal}, Andre and {Hessel}, Kaitlyn and {Marois}, Christian and {Rochon}, Alexandra},
        title = "{Dynamical Mass Constraints on Transition Disk Perturbers with the G23H Catalog}",
      journal = {arXiv e-prints},
         year = 2026,
        month = feb,
          eid = {arXiv:2602.07731},
        pages = {arXiv:2602.07731},
          doi = {10.48550/arXiv.2602.07731},
archivePrefix = {arXiv},
       eprint = {2602.07731},
 primaryClass = {astro-ph.EP},
       adsurl = {https://ui.adsabs.harvard.edu/abs/2026arXiv260207731B}
}

@ARTICLE{Dong2015a,
       author = {{Dong}, Ruobing and {Zhu}, Zhaohuan and {Rafikov}, Roman R. and {Stone}, James M.},
        title = "{Observational Signatures of Planets in Protoplanetary Disks: Spiral Arms Observed in Scattered Light Imaging Can be Induced by Planets}",
      journal = {\apjl},
         year = 2015,
        month = aug,
       volume = {809},
       number = {1},
          eid = {L5},
        pages = {L5},
          doi = {10.1088/2041-8205/809/1/L5},
archivePrefix = {arXiv},
       eprint = {1507.03596},
 primaryClass = {astro-ph.EP},
       adsurl = {https://ui.adsabs.harvard.edu/abs/2015ApJ...809L...5D}
}

@ARTICLE{Dong2015b,
       author = {{Dong}, Ruobing and {Zhu}, Zhaohuan and {Whitney}, Barbara},
        title = "{Observational Signatures of Planets in Protoplanetary Disks I. Gaps Opened by Single and Multiple Young Planets in Disks}",
      journal = {\apj},
         year = 2015,
        month = aug,
       volume = {809},
       number = {1},
          eid = {93},
        pages = {93},
          doi = {10.1088/0004-637X/809/1/93},
archivePrefix = {arXiv},
       eprint = {1411.6063},
 primaryClass = {astro-ph.EP},
       adsurl = {https://ui.adsabs.harvard.edu/abs/2015ApJ...809...93D}
}

@ARTICLE{Smallwood2019,
       author = {{Smallwood}, Jeremy L. and {Lubow}, Stephen H. and {Franchini}, Alessia and {Martin}, Rebecca G.},
        title = "{Alignment of a circumbinary disc around an eccentric binary with application to KH 15D}",
      journal = {\mnras},
         year = 2019,
        month = jun,
       volume = {486},
       number = {2},
        pages = {2919-2932},
          doi = {10.1093/mnras/stz994},
archivePrefix = {arXiv},
       eprint = {1904.00985},
 primaryClass = {astro-ph.EP},
       adsurl = {https://ui.adsabs.harvard.edu/abs/2019MNRAS.486.2919S}
}

@ARTICLE{Lubow2018,
       author = {{Lubow}, Stephen H. and {Martin}, Rebecca G.},
        title = "{Linear analysis of the evolution of nearly polar low-mass circumbinary discs}",
      journal = {\mnras},
         year = 2018,
        month = jan,
       volume = {473},
       number = {3},
        pages = {3733-3746},
          doi = {10.1093/mnras/stx2643},
archivePrefix = {arXiv},
       eprint = {1710.02233},
 primaryClass = {astro-ph.SR},
       adsurl = {https://ui.adsabs.harvard.edu/abs/2018MNRAS.473.3733L}
}

@ARTICLE{Shakura1973,
       author = {{Shakura}, N.~I. and {Sunyaev}, R.~A.},
        title = "{Black holes in binary systems. Observational appearance.}",
      journal = {\aap},
         year = 1973,
        month = jan,
       volume = {24},
        pages = {337-355},
       adsurl = {https://ui.adsabs.harvard.edu/abs/1973A&A....24..337S}
}

@ARTICLE{Facchini2018,
       author = {{Facchini}, Stefano and {Juh{\'a}sz}, Attila and {Lodato}, Giuseppe},
        title = "{Signatures of broken protoplanetary discs in scattered light and in sub-millimetre observations}",
      journal = {\mnras},
         year = 2018,
        month = feb,
       volume = {473},
       number = {4},
        pages = {4459-4475},
          doi = {10.1093/mnras/stx2523},
archivePrefix = {arXiv},
       eprint = {1709.08369},
 primaryClass = {astro-ph.SR},
       adsurl = {https://ui.adsabs.harvard.edu/abs/2018MNRAS.473.4459F}
}

@ARTICLE{Lubow2000,
       author = {{Lubow}, S.~H. and {Ogilvie}, G.~I.},
        title = "{On the Tilting of Protostellar Disks by Resonant Tidal Effects}",
      journal = {\apj},
         year = 2000,
        month = jul,
       volume = {538},
       number = {1},
        pages = {326-340},
          doi = {10.1086/309101},
archivePrefix = {arXiv},
       eprint = {astro-ph/0003028},
 primaryClass = {astro-ph},
       adsurl = {https://ui.adsabs.harvard.edu/abs/2000ApJ...538..326L}
}

@ARTICLE{Kimmig2026,
       author = {{Kimmig}, C.~N. and {Weber}, P. and {Rosotti}, G.~P. and {Facchini}, S. and {Dullemond}, C.~P.},
        title = "{Warps survive beyond fly-by encounters in protoplanetary disks. RW Aur A as a case study}",
      journal = {arXiv e-prints},
         year = 2026,
        month = jan,
          eid = {arXiv:2601.22223},
        pages = {arXiv:2601.22223},
          doi = {10.48550/arXiv.2601.22223},
archivePrefix = {arXiv},
       eprint = {2601.22223},
 primaryClass = {astro-ph.EP},
       adsurl = {https://ui.adsabs.harvard.edu/abs/2026arXiv260122223K}
}

@ARTICLE{Garufi2024,
       author = {{Garufi}, A. and {Ginski}, C. and {van Holstein}, R.~G. and {Benisty}, M. and {Manara}, C.~F. and {P{\'e}rez}, S. and {Pinilla}, P. and {Ribas}, {\'A}. and {Weber}, P. and {Williams}, J. and {Cieza}, L. and {Dominik}, C. and {Facchini}, S. and {Huang}, J. and {Zurlo}, A. and {Bae}, J. and {Hagelberg}, J. and {Henning}, Th. and {Hogerheijde}, M.~R. and {Janson}, M. and {M{\'e}nard}, F. and {Messina}, S. and {Meyer}, M.~R. and {Pinte}, C. and {Quanz}, S.~P. and {Rigliaco}, E. and {Roccatagliata}, V. and {Schmid}, H.~M. and {Szul{\'a}gyi}, J. and {van Boekel}, R. and {Wahhaj}, Z. and {Antichi}, J. and {Baruffolo}, A. and {Moulin}, T.},
        title = "{The SPHERE view of the Taurus star-forming region. The full census of planet-forming disks with GTO and DESTINYS programs}",
      journal = {\aap},
         year = 2024,
        month = may,
       volume = {685},
          eid = {A53},
        pages = {A53},
          doi = {10.1051/0004-6361/202347586},
archivePrefix = {arXiv},
       eprint = {2403.02158},
 primaryClass = {astro-ph.GA},
       adsurl = {https://ui.adsabs.harvard.edu/abs/2024A&A...685A..53G}
}

@ARTICLE{Wallack2024,
       author = {{Wallack}, Nicole L. and {Ruffio}, Jean-Baptiste and {Ruane}, Garreth and {Ren}, Bin B. and {Xuan}, Jerry W. and {Villenave}, Marion and {Mawet}, Dimitri and {Stapelfeldt}, Karl and {Wang}, Jason J. and {Liu}, Michael C. and {Absil}, Olivier and {Alvarez}, Carlos and {Bae}, Jaehan and {Bond}, Charlotte and {Bottom}, Michael and {Calvin}, Benjamin and {Choquet}, {\'E}lodie and {Christiaens}, Valentin and {Cook}, Therese and {Femen{\'\i}a Castell{\'a}}, Bruno and {Gomez Gonzalez}, Carlos and {Guidi}, Greta and {Huby}, Elsa and {Kastner}, Joel and {Knutson}, Heather A. and {Meshkat}, Tiffany and {Ngo}, Henry and {Ragland}, Sam and {Reggiani}, Maddalena and {Ricci}, Luca and {Serabyn}, Eugene and {Uyama}, Taichi and {Williams}, Jonathan P. and {Wizinowich}, Peter and {Zawol}, Zoe and {Zhang}, Shangjia and {Zhu}, Zhaohuan},
        title = "{A Survey of Protoplanetary Disks Using the Keck/NIRC2 Vortex Coronagraph}",
      journal = {\aj},
         year = 2024,
        month = aug,
       volume = {168},
       number = {2},
          eid = {78},
        pages = {78},
          doi = {10.3847/1538-3881/ad390c},
archivePrefix = {arXiv},
       eprint = {2408.04048},
 primaryClass = {astro-ph.EP},
       adsurl = {https://ui.adsabs.harvard.edu/abs/2024AJ....168...78W}
}

@ARTICLE{Muto2012,
       author = {{Muto}, T. and {Grady}, C.~A. and {Hashimoto}, J. and {Fukagawa}, M. and {Hornbeck}, J.~B. and {Sitko}, M. and {Russell}, R. and {Werren}, C. and {Cur{\'e}}, M. and {Currie}, T. and {Ohashi}, N. and {Okamoto}, Y. and {Momose}, M. and {Honda}, M. and {Inutsuka}, S. and {Takeuchi}, T. and {Dong}, R. and {Abe}, L. and {Brandner}, W. and {Brandt}, T. and {Carson}, J. and {Egner}, S. and {Feldt}, M. and {Fukue}, T. and {Goto}, M. and {Guyon}, O. and {Hayano}, Y. and {Hayashi}, M. and {Hayashi}, S. and {Henning}, T. and {Hodapp}, K.~W. and {Ishii}, M. and {Iye}, M. and {Janson}, M. and {Kandori}, R. and {Knapp}, G.~R. and {Kudo}, T. and {Kusakabe}, N. and {Kuzuhara}, M. and {Matsuo}, T. and {Mayama}, S. and {McElwain}, M.~W. and {Miyama}, S. and {Morino}, J.-I. and {Moro-Martin}, A. and {Nishimura}, T. and {Pyo}, T.-S. and {Serabyn}, E. and {Suto}, H. and {Suzuki}, R. and {Takami}, M. and {Takato}, N. and {Terada}, H. and {Thalmann}, C. and {Tomono}, D. and {Turner}, E.~L. and {Watanabe}, M. and {Wisniewski}, J.~P. and {Yamada}, T. and {Takami}, H. and {Usuda}, T. and {Tamura}, M.},
        title = "{Discovery of Small-scale Spiral Structures in the Disk of SAO 206462 (HD 135344B): Implications for the Physical State of the Disk from Spiral Density Wave Theory}",
      journal = {\apjl},
         year = 2012,
        month = apr,
       volume = {748},
       number = {2},
          eid = {L22},
        pages = {L22},
          doi = {10.1088/2041-8205/748/2/L22},
archivePrefix = {arXiv},
       eprint = {1202.6139},
 primaryClass = {astro-ph.EP},
       adsurl = {https://ui.adsabs.harvard.edu/abs/2012ApJ...748L..22M}
}

@ARTICLE{Kuo2022,
       author = {{Kuo}, I.-Hsuan Genevieve and {Yen}, Hsi-Wei and {Gu}, Pin-Gao and {Chang}, Tze-En},
        title = "{Kinematical Constraint on Eccentricity in the Protoplanetary Disk MWC 758 with ALMA}",
      journal = {\apj},
         year = 2022,
        month = oct,
       volume = {938},
       number = {1},
          eid = {50},
        pages = {50},
          doi = {10.3847/1538-4357/ac9228},
archivePrefix = {arXiv},
       eprint = {2209.12741},
 primaryClass = {astro-ph.EP},
       adsurl = {https://ui.adsabs.harvard.edu/abs/2022ApJ...938...50K}
}

@ARTICLE{Long2018,
       author = {{Long}, Feng and {Pinilla}, Paola and {Herczeg}, Gregory J. and {Harsono}, Daniel and {Dipierro}, Giovanni and {Pascucci}, Ilaria and {Hendler}, Nathan and {Tazzari}, Marco and {Ragusa}, Enrico and {Salyk}, Colette and {Edwards}, Suzan and {Lodato}, Giuseppe and {van de Plas}, Gerrit and {Johnstone}, Doug and {Liu}, Yao and {Boehler}, Yann and {Cabrit}, Sylvie and {Manara}, Carlo F. and {Menard}, Francois and {Mulders}, Gijs D. and {Nisini}, Brunella and {Fischer}, William J. and {Rigliaco}, Elisabetta and {Banzatti}, Andrea and {Avenhaus}, Henning and {Gully-Santiago}, Michael},
        title = "{Gaps and Rings in an ALMA Survey of Disks in the Taurus Star-forming Region}",
      journal = {\apj},
         year = 2018,
        month = dec,
       volume = {869},
       number = {1},
          eid = {17},
        pages = {17},
          doi = {10.3847/1538-4357/aae8e1},
archivePrefix = {arXiv},
       eprint = {1810.06044},
 primaryClass = {astro-ph.SR},
       adsurl = {https://ui.adsabs.harvard.edu/abs/2018ApJ...869...17L}
}

@ARTICLE{Andrews2018,
       author = {{Andrews}, Sean M. and {Huang}, Jane and {P{\'e}rez}, Laura M. and {Isella}, Andrea and {Dullemond}, Cornelis P. and {Kurtovic}, Nicol{\'a}s T. and {Guzm{\'a}n}, Viviana V. and {Carpenter}, John M. and {Wilner}, David J. and {Zhang}, Shangjia and {Zhu}, Zhaohuan and {Birnstiel}, Tilman and {Bai}, Xue-Ning and {Benisty}, Myriam and {Hughes}, A. Meredith and {{\"O}berg}, Karin I. and {Ricci}, Luca},
        title = "{The Disk Substructures at High Angular Resolution Project (DSHARP). I. Motivation, Sample, Calibration, and Overview}",
      journal = {\apjl},
         year = 2018,
        month = dec,
       volume = {869},
       number = {2},
          eid = {L41},
        pages = {L41},
          doi = {10.3847/2041-8213/aaf741},
archivePrefix = {arXiv},
       eprint = {1812.04040},
 primaryClass = {astro-ph.SR},
       adsurl = {https://ui.adsabs.harvard.edu/abs/2018ApJ...869L..41A}
}

@ARTICLE{Juhasz2017,
       author = {{Juh{\'a}sz}, Attila and {Facchini}, Stefano},
        title = "{Observational signatures of linear warps in circumbinary discs}",
      journal = {\mnras},
         year = 2017,
        month = apr,
       volume = {466},
       number = {4},
        pages = {4053-4073},
          doi = {10.1093/mnras/stw3389},
archivePrefix = {arXiv},
       eprint = {1701.02611},
 primaryClass = {astro-ph.EP},
       adsurl = {https://ui.adsabs.harvard.edu/abs/2017MNRAS.466.4053J}
}

@ARTICLE{Foucart2011,
       author = {{Foucart}, Francois and {Lai}, Dong},
        title = "{Evolution of spin direction of accreting magnetic protostars and spin-orbit misalignment in exoplanetary systems - II. Warped discs}",
      journal = {\mnras},
         year = 2011,
        month = apr,
       volume = {412},
       number = {4},
        pages = {2799-2815},
          doi = {10.1111/j.1365-2966.2010.18176.x},
archivePrefix = {arXiv},
       eprint = {1009.3233},
 primaryClass = {astro-ph.EP},
       adsurl = {https://ui.adsabs.harvard.edu/abs/2011MNRAS.412.2799F}
}

@ARTICLE{Bate2010,
       author = {{Bate}, M.~R. and {Lodato}, G. and {Pringle}, J.~E.},
        title = "{Chaotic star formation and the alignment of stellar rotation with disc and planetary orbital axes}",
      journal = {\mnras},
         year = 2010,
        month = jan,
       volume = {401},
       number = {3},
        pages = {1505-1513},
          doi = {10.1111/j.1365-2966.2009.15773.x},
archivePrefix = {arXiv},
       eprint = {0909.4255},
 primaryClass = {astro-ph.SR},
       adsurl = {https://ui.adsabs.harvard.edu/abs/2010MNRAS.401.1505B}
}

@ARTICLE{Kraus2020,
       author = {{Kraus}, Stefan and {Kreplin}, Alexander and {Young}, Alison K. and {Bate}, Matthew R. and {Monnier}, John D. and {Harries}, Tim J. and {Avenhaus}, Henning and {Kluska}, Jacques and {Laws}, Anna S.~E. and {Rich}, Evan A. and {Willson}, Matthew and {Aarnio}, Alicia N. and {Adams}, Fred C. and {Andrews}, Sean M. and {Anugu}, Narsireddy and {Bae}, Jaehan and {ten Brummelaar}, Theo and {Calvet}, Nuria and {Cur{\'e}}, Michel and {Davies}, Claire L. and {Ennis}, Jacob and {Espaillat}, Catherine and {Gardner}, Tyler and {Hartmann}, Lee and {Hinkley}, Sasha and {Labdon}, Aaron and {Lanthermann}, Cyprien and {LeBouquin}, Jean-Baptiste and {Schaefer}, Gail H. and {Setterholm}, Benjamin R. and {Wilner}, David and {Zhu}, Zhaohuan},
        title = "{A triple-star system with a misaligned and warped circumstellar disk shaped by disk tearing}",
      journal = {Science},
         year = 2020,
        month = sep,
       volume = {369},
       number = {6508},
        pages = {1233-1238},
          doi = {10.1126/science.aba4633},
archivePrefix = {arXiv},
       eprint = {2004.01204},
 primaryClass = {astro-ph.SR},
       adsurl = {https://ui.adsabs.harvard.edu/abs/2020Sci...369.1233K}
}

@ARTICLE{Zhu2019,
       author = {{Zhu}, Zhaohuan},
        title = "{Inclined massive planets in a protoplanetary disc: gap opening, disc breaking, and observational signatures}",
      journal = {\mnras},
         year = 2019,
        month = mar,
       volume = {483},
       number = {3},
        pages = {4221-4241},
          doi = {10.1093/mnras/sty3358},
archivePrefix = {arXiv},
       eprint = {1812.01262},
 primaryClass = {astro-ph.EP},
       adsurl = {https://ui.adsabs.harvard.edu/abs/2019MNRAS.483.4221Z}
}

@ARTICLE{Lai1999,
       author = {{Lai}, Dong},
        title = "{Magnetically Driven Warping, Precession, and Resonances in Accretion Disks}",
      journal = {\apj},
         year = 1999,
        month = oct,
       volume = {524},
       number = {2},
        pages = {1030-1047},
          doi = {10.1086/307850},
archivePrefix = {arXiv},
       eprint = {astro-ph/9904110},
 primaryClass = {astro-ph},
       adsurl = {https://ui.adsabs.harvard.edu/abs/1999ApJ...524.1030L}
}

@ARTICLE{Ginski2021,
       author = {{Ginski}, Christian and {Facchini}, Stefano and {Huang}, Jane and {Benisty}, Myriam and {Vaendel}, Dennis and {Stapper}, Lucas and {Dominik}, Carsten and {Bae}, Jaehan and {M{\'e}nard}, Fran{\c{c}}ois and {Muro-Arena}, Gabriela and {Hogerheijde}, Michiel R. and {McClure}, Melissa and {van Holstein}, Rob G. and {Birnstiel}, Tilman and {Boehler}, Yann and {Bohn}, Alexander and {Flock}, Mario and {Mamajek}, Eric E. and {Manara}, Carlo F. and {Pinilla}, Paola and {Pinte}, Christophe and {Ribas}, {\'A}lvaro},
        title = "{Disk Evolution Study Through Imaging of Nearby Young Stars (DESTINYS): Late Infall Causing Disk Misalignment and Dynamic Structures in SU Aur}",
      journal = {\apjl},
         year = 2021,
        month = feb,
       volume = {908},
       number = {2},
          eid = {L25},
        pages = {L25},
          doi = {10.3847/2041-8213/abdf57},
archivePrefix = {arXiv},
       eprint = {2102.08781},
 primaryClass = {astro-ph.EP},
       adsurl = {https://ui.adsabs.harvard.edu/abs/2021ApJ...908L..25G}
}

@ARTICLE{Tanious2024,
       author = {{Tanious}, M. and {Le Gal}, R. and {Neri}, R. and {Faure}, A. and {Gupta}, A. and {Law}, C.~J. and {Huang}, J. and {Cuello}, N. and {Williams}, J.~P. and {M{\'e}nard}, F.},
        title = "{Anatomy of the Class I protostar L1489 IRS with NOEMA. I. Disk, streamers, outflow(s) and bubbles at 3 mm}",
      journal = {\aap},
         year = 2024,
        month = jul,
       volume = {687},
          eid = {A92},
        pages = {A92},
          doi = {10.1051/0004-6361/202348785},
archivePrefix = {arXiv},
       eprint = {2403.18905},
 primaryClass = {astro-ph.SR},
       adsurl = {https://ui.adsabs.harvard.edu/abs/2024A&A...687A..92T}
}

@ARTICLE{Sai2020,
       author = {{Sai}, Jinshi and {Ohashi}, Nagayoshi and {Saigo}, Kazuya and {Matsumoto}, Tomoaki and {Aso}, Yusuke and {Takakuwa}, Shigehisa and {Aikawa}, Yuri and {Kurose}, Ippei and {Yen}, Hsi-Wei and {Tomisaka}, Kohji and {Tomida}, Kengo and {Machida}, Masahiro N.},
        title = "{Disk Structure around the Class I Protostar L1489 IRS Revealed by ALMA: A Warped-disk System}",
      journal = {\apj},
         year = 2020,
        month = apr,
       volume = {893},
       number = {1},
          eid = {51},
        pages = {51},
          doi = {10.3847/1538-4357/ab8065},
archivePrefix = {arXiv},
       eprint = {2003.07067},
 primaryClass = {astro-ph.SR},
       adsurl = {https://ui.adsabs.harvard.edu/abs/2020ApJ...893...51S}
}

@ARTICLE{Dong2018,
       author = {{Dong}, Ruobing and {Liu}, Sheng-yuan and {Eisner}, Josh and {Andrews}, Sean and {Fung}, Jeffrey and {Zhu}, Zhaohuan and {Chiang}, Eugene and {Hashimoto}, Jun and {Liu}, Hauyu Baobab and {Casassus}, Simon and {Esposito}, Thomas and {Hasegawa}, Yasuhiro and {Muto}, Takayuki and {Pavlyuchenkov}, Yaroslav and {Wilner}, David and {Akiyama}, Eiji and {Tamura}, Motohide and {Wisniewski}, John},
        title = "{The Eccentric Cavity, Triple Rings, Two-armed Spirals, and Double Clumps of the MWC 758 Disk}",
      journal = {\apj},
         year = 2018,
        month = jun,
       volume = {860},
       number = {2},
          eid = {124},
        pages = {124},
          doi = {10.3847/1538-4357/aac6cb},
archivePrefix = {arXiv},
       eprint = {1805.12141},
 primaryClass = {astro-ph.SR},
       adsurl = {https://ui.adsabs.harvard.edu/abs/2018ApJ...860..124D}
}

@ARTICLE{Zhang2025,
       author = {{Zhang}, Shangjia and {Zhu}, Zhaohuan and {Fairbairn}, Callum W.},
        title = "{Shadow-induced Warps in Protoplanetary Disks}",
      journal = {\apjl},
         year = 2025,
        month = dec,
       volume = {995},
       number = {1},
          eid = {L33},
        pages = {L33},
          doi = {10.3847/2041-8213/ae2023},
archivePrefix = {arXiv},
       eprint = {2511.11358},
 primaryClass = {astro-ph.EP},
       adsurl = {https://ui.adsabs.harvard.edu/abs/2025ApJ...995L..33Z}
}

@ARTICLE{Bi2020,
       author = {{Bi}, Jiaqing and {van der Marel}, Nienke and {Dong}, Ruobing and {Muto}, Takayuki and {Martin}, Rebecca G. and {Smallwood}, Jeremy L. and {Hashimoto}, Jun and {Liu}, Hauyu Baobab and {Nomura}, Hideko and {Hasegawa}, Yasuhiro and {Takami}, Michihiro and {Konishi}, Mihoko and {Momose}, Munetake and {Kanagawa}, Kazuhiro D. and {Kataoka}, Akimasa and {Ono}, Tomohiro and {Sitko}, Michael L. and {Takahashi}, Sanemichi Z. and {Tomida}, Kengo and {Tsukagoshi}, Takashi},
        title = "{GW Ori: Interactions between a Triple-star System and Its Circumtriple Disk in Action}",
      journal = {\apjl},
         year = 2020,
        month = may,
       volume = {895},
       number = {1},
          eid = {L18},
        pages = {L18},
          doi = {10.3847/2041-8213/ab8eb4},
archivePrefix = {arXiv},
       eprint = {2004.03135},
 primaryClass = {astro-ph.SR},
       adsurl = {https://ui.adsabs.harvard.edu/abs/2020ApJ...895L..18B}
}

@article{Winter2025,
doi = {10.3847/2041-8213/adf113},
url = {https://doi.org/10.3847/2041-8213/adf113},
year = {2025},
month = {aug},
publisher = {The American Astronomical Society},
volume = {990},
number = {1},
pages = {L10},
author = {Winter, Andrew J. and Benisty, Myriam and Izquierdo, Andrés F. and Lodato, Giuseppe and Teague, Richard and Kimmig, Carolin N. and Andrews, Sean M. and Bae, Jaehan and Barraza-Alfaro, Marcelo and Cuello, Nicolás and Curone, Pietro and Czekala, Ian and Facchini, Stefano and Fasano, Daniele and Hall, Cassandra and Hardiman, Caitlyn and Hilder, Thomas and Ilee, John D. and Fukagawa, Misato and Longarini, Cristiano and Ménard, François and Orihara, Ryuta and Pinte, Christophe and Price, Daniel J. and Rosotti, Giovanni and Stadler, Jochen and Wilner, David J. and Wölfer, Lisa and Yen, Hsi-Wei and Yoshida, Tomohiro C. and Zawadzki, Brianna},
title = {exoALMA. XVIII. Interpreting Large-scale Kinematic Structures as Moderate Warping},
journal = {The Astrophysical Journal Letters}
}

@ARTICLE{Ren2023,
       author = {{Ren}, Bin B. and {Benisty}, Myriam and {Ginski}, Christian and {Tazaki}, Ryo and {Wallack}, Nicole L. and {Milli}, Julien and {Garufi}, Antonio and {Bae}, Jaehan and {Facchini}, Stefano and {M{\'e}nard}, Fran{\c{c}}ois and {Pinilla}, Paola and {Swastik}, C. and {Teague}, Richard and {Wahhaj}, Zahed},
        title = "{Protoplanetary disks in K$_{s}$-band total intensity and polarized light}",
      journal = {\aap},
         year = 2023,
        month = dec,
       volume = {680},
          eid = {A114},
        pages = {A114},
          doi = {10.1051/0004-6361/202347353},
archivePrefix = {arXiv},
       eprint = {2310.08589},
 primaryClass = {astro-ph.EP},
       adsurl = {https://ui.adsabs.harvard.edu/abs/2023A&A...680A.114R}
}

@ARTICLE{Hammond2022,
       author = {{Hammond}, Iain and {Christiaens}, Valentin and {Price}, Daniel J. and {Ubeira-Gabellini}, Maria Giulia and {Baird}, Jennifer and {Calcino}, Josh and {Benisty}, Myriam and {Lodato}, Giuseppe and {Testi}, Leonardo and {Pinte}, Christophe and {Toci}, Claudia and {Fedele}, Davide},
        title = "{External or internal companion exciting the spiral arms in CQ Tau?}",
      journal = {\mnras},
         year = 2022,
        month = oct,
       volume = {515},
       number = {4},
        pages = {6109-6121},
          doi = {10.1093/mnras/stac2119},
archivePrefix = {arXiv},
       eprint = {2207.08587},
 primaryClass = {astro-ph.EP},
       adsurl = {https://ui.adsabs.harvard.edu/abs/2022MNRAS.515.6109H}
}

@ARTICLE{Bohn2022,
       author = {{Bohn}, A.~J. and {Benisty}, M. and {Perraut}, K. and {van der Marel}, N. and {W{\"o}lfer}, L. and {van Dishoeck}, E.~F. and {Facchini}, S. and {Manara}, C.~F. and {Teague}, R. and {Francis}, L. and {Berger}, J.-P. and {Garcia-Lopez}, R. and {Ginski}, C. and {Henning}, T. and {Kenworthy}, M. and {Kraus}, S. and {M{\'e}nard}, F. and {M{\'e}rand}, A. and {P{\'e}rez}, L.~M.},
        title = "{Probing inner and outer disk misalignments in transition disks. Constraints from VLTI/GRAVITY and ALMA observations}",
      journal = {\aap},
         year = 2022,
        month = feb,
       volume = {658},
          eid = {A183},
        pages = {A183},
          doi = {10.1051/0004-6361/202142070},
archivePrefix = {arXiv},
       eprint = {2112.00123},
 primaryClass = {astro-ph.EP},
       adsurl = {https://ui.adsabs.harvard.edu/abs/2022A&A...658A.183B}
}

@ARTICLE{Zagaria2025,
       author = {{Zagaria}, Francesco and {Jiang}, Haochang and {Cataldi}, Gianni and {Facchini}, Stefano and {Benisty}, Myriam and {Aikawa}, Yuri and {Andrews}, Sean and {Bae}, Jaehan and {Barraza-Alfaro}, Marcelo and {Curone}, Pietro and {Czekala}, Ian and {Fasano}, Daniele and {Hall}, Cassandra and {Hammond}, Iain and {Huang}, Jane and {Ilee}, John D. and {Izquierdo}, Andr{\'e}s F. and {Lawrence}, Jensen and {Lodato}, Giuseppe and {M{\'e}nard}, Fran{\c{c}}ois and {Pinte}, Christophe and {Rosotti}, Giovanni P. and {Stadler}, Jochen and {Teague}, Richard and {Testi}, Leonardo and {Wilner}, David and {Winter}, Andrew and {Yoshida}, Tomohiro},
        title = "{SO Emission in the Dynamically Perturbed Protoplanetary Disks around CQ Tau and MWC 758}",
      journal = {\apj},
         year = 2025,
        month = aug,
       volume = {989},
       number = {1},
          eid = {30},
        pages = {30},
          doi = {10.3847/1538-4357/ade683},
archivePrefix = {arXiv},
       eprint = {2506.16481},
 primaryClass = {astro-ph.EP},
       adsurl = {https://ui.adsabs.harvard.edu/abs/2025ApJ...989...30Z}
}

@ARTICLE{Curone2025,
       author = {{Curone}, Pietro and {Facchini}, Stefano and {Andrews}, Sean M. and {Testi}, Leonardo and {Benisty}, Myriam and {Czekala}, Ian and {Huang}, Jane and {Ilee}, John D. and {Isella}, Andrea and {Lodato}, Giuseppe and {Loomis}, Ryan A. and {Stadler}, Jochen and {Winter}, Andrew J. and {Bae}, Jaehan and {Barraza-Alfaro}, Marcelo and {Cataldi}, Gianni and {Cuello}, Nicol{\'a}s and {Fasano}, Daniele and {Flock}, Mario and {Fukagawa}, Misato and {Galloway-Sprietsma}, Maria and {Garg}, Himanshi and {Hall}, Cassandra and {Izquierdo}, Andr{\'e}s F. and {Kanagawa}, Kazuhiro and {Lesur}, Geoffroy and {Longarini}, Cristiano and {Menard}, Francois and {Orihara}, Ryuta and {Pinte}, Christophe and {Price}, Daniel J. and {Rosotti}, Giovanni and {Teague}, Richard and {Wafflard-Fernandez}, Gaylor and {Wilner}, David J. and {W{\"o}lfer}, Lisa and {Yen}, Hsi-Wei and {Yoshida}, Tomohiro C. and {Zawadzki}, Brianna},
        title = "{exoALMA. IV. Substructures, Asymmetries, and the Faint Outer Disk in Continuum Emission}",
      journal = {\apjl},
         year = 2025,
        month = may,
       volume = {984},
       number = {1},
          eid = {L9},
        pages = {L9},
          doi = {10.3847/2041-8213/adc438},
archivePrefix = {arXiv},
       eprint = {2504.18725},
 primaryClass = {astro-ph.EP},
       adsurl = {https://ui.adsabs.harvard.edu/abs/2025ApJ...984L...9C}
}

@software{Nakazato2020,
       author = {{Nakazato}, Takeshi and {Ikeda}, Shiro},
        title = "{PRIISM: Python module for Radio Interferometry Imaging with Sparse Modeling}",
 howpublished = {Astrophysics Source Code Library, record ascl:2006.002},
         year = 2020,
        month = jun,
          eid = {ascl:2006.002},
archivePrefix = {ascl},
       eprint = {2006.002},
       adsurl = {https://ui.adsabs.harvard.edu/abs/2020ascl.soft06002N}
}

@ARTICLE{Shoshi2025b,
       author = {{Shoshi}, Ayumu and {Yamaguchi}, Masayuki and {Muto}, Takayuki and {Hirano}, Naomi and {Kawabe}, Ryohei and {Tsukagoshi}, Takashi and {Machida}, Masahiro N.},
        title = "{ALMA 2D super-resolution imaging survey of Ophiuchus Class I/flat spectrum/II disks. I. Discovery of new disk substructures}",
      journal = {\pasj},
         year = 2025,
        month = jun,
       volume = {77},
       number = {3},
        pages = {572-606},
          doi = {10.1093/pasj/psaf026},
archivePrefix = {arXiv},
       eprint = {2503.21060},
 primaryClass = {astro-ph.EP},
       adsurl = {https://ui.adsabs.harvard.edu/abs/2025PASJ...77..572S}
}

@ARTICLE{Shoshi2025a,
       author = {{Shoshi}, Ayumu and {Muto}, Takayuki and {Bosschaart}, Quincy and {van der Marel}, Nienke and {Mulders}, Gijs D. and {Omura}, Mitsuki and {Tokuda}, Kazuki and {Machida}, Masahiro N.},
        title = "{ALMA High-resolution Observation for the Transitional Disk Around IRAS 04125+2902}",
      journal = {\apj},
         year = 2025,
        month = nov,
       volume = {993},
       number = {1},
          eid = {90},
        pages = {90},
          doi = {10.3847/1538-4357/ae045a},
archivePrefix = {arXiv},
       eprint = {2509.01896},
 primaryClass = {astro-ph.EP},
       adsurl = {https://ui.adsabs.harvard.edu/abs/2025ApJ...993...90S}
}

@ARTICLE{Yamaguchi2021,
       author = {{Yamaguchi}, Masayuki and {Tsukagoshi}, Takashi and {Muto}, Takayuki and {Nomura}, Hideko and {Nakazato}, Takeshi and {Ikeda}, Shiro and {Tamura}, Motohide and {Kawabe}, Ryohei},
        title = "{ALMA Super-resolution Imaging of T Tau: r = 12 au Gap in the Compact Dust Disk around T Tau N}",
      journal = {\apj},
         year = 2021,
        month = dec,
       volume = {923},
       number = {1},
          eid = {121},
        pages = {121},
          doi = {10.3847/1538-4357/ac2bfd},
archivePrefix = {arXiv},
       eprint = {2110.00974},
 primaryClass = {astro-ph.EP},
       adsurl = {https://ui.adsabs.harvard.edu/abs/2021ApJ...923..121Y}
}

@ARTICLE{Yamaguchi2020,
       author = {{Yamaguchi}, Masayuki and {Akiyama}, Kazunori and {Tsukagoshi}, Takashi and {Muto}, Takayuki and {Kataoka}, Akimasa and {Tazaki}, Fumie and {Ikeda}, Shiro and {Fukagawa}, Misato and {Honma}, Mareki and {Kawabe}, Ryohei},
        title = "{Super-resolution Imaging of the Protoplanetary Disk HD 142527 Using Sparse Modeling}",
      journal = {\apj},
         year = 2020,
        month = jun,
       volume = {895},
       number = {2},
          eid = {84},
        pages = {84},
          doi = {10.3847/1538-4357/ab899f},
archivePrefix = {arXiv},
       eprint = {2004.10989},
 primaryClass = {astro-ph.EP},
       adsurl = {https://ui.adsabs.harvard.edu/abs/2020ApJ...895...84Y}
}

@INPROCEEDINGS{McMullin2007,
       author = {{McMullin}, J.~P. and {Waters}, B. and {Schiebel}, D. and {Young}, W. and {Golap}, K.},
        title = "{CASA Architecture and Applications}",
    booktitle = {Astronomical Data Analysis Software and Systems XVI},
         year = 2007,
       editor = {{Shaw}, R.~A. and {Hill}, F. and {Bell}, D.~J.},
       series = {Astronomical Society of the Pacific Conference Series},
       volume = {376},
        month = oct,
        pages = {127},
       adsurl = {https://ui.adsabs.harvard.edu/abs/2007ASPC..376..127M}
}

@ARTICLE{Wolfer2021,
       author = {{W{\"o}lfer}, L. and {Facchini}, S. and {Kurtovic}, N.~T. and {Teague}, R. and {van Dishoeck}, E.~F. and {Benisty}, M. and {Ercolano}, B. and {Lodato}, G. and {Miotello}, A. and {Rosotti}, G. and {Testi}, L. and {Ubeira Gabellini}, M.~G.},
        title = "{A highly non-Keplerian protoplanetary disc. Spiral structure in the gas disc of CQ Tau}",
      journal = {\aap},
         year = 2021,
        month = apr,
       volume = {648},
          eid = {A19},
        pages = {A19},
          doi = {10.1051/0004-6361/202039469},
archivePrefix = {arXiv},
       eprint = {2012.04680},
 primaryClass = {astro-ph.EP},
       adsurl = {https://ui.adsabs.harvard.edu/abs/2021A&A...648A..19W}
}

@ARTICLE{UbeiraGabellini2019,
       author = {{Ubeira Gabellini}, M. Giulia and {Miotello}, Anna and {Facchini}, Stefano and {Ragusa}, Enrico and {Lodato}, Giuseppe and {Testi}, Leonardo and {Benisty}, Myriam and {Bruderer}, Simon and {T. Kurtovic}, Nicol{\'a}s and {Andrews}, Sean and {Carpenter}, John and {Corder}, Stuartt A. and {Dipierro}, Giovanni and {Ercolano}, Barbara and {Fedele}, Davide and {Guidi}, Greta and {Henning}, Thomas and {Isella}, Andrea and {Kwon}, Woojin and {Linz}, Hendrik and {McClure}, Melissa and {Perez}, Laura and {Ricci}, Luca and {Rosotti}, Giovanni and {Tazzari}, Marco and {Wilner}, David},
        title = "{A dust and gas cavity in the disc around CQ Tau revealed by ALMA}",
      journal = {\mnras},
         year = 2019,
        month = jul,
       volume = {486},
       number = {4},
        pages = {4638-4654},
          doi = {10.1093/mnras/stz1138},
archivePrefix = {arXiv},
       eprint = {1905.00909},
 primaryClass = {astro-ph.EP},
       adsurl = {https://ui.adsabs.harvard.edu/abs/2019MNRAS.486.4638U}
}

@ARTICLE{Yamaguchi2024,
       author = {{Yamaguchi}, Masayuki and {Muto}, Takayuki and {Tsukagoshi}, Takashi and {Nomura}, Hideko and {Hirano}, Naomi and {Nakazato}, Takeshi and {Ikeda}, Shiro and {Tamura}, Motohide and {Kawabe}, Ryohei},
        title = "{ALMA 2D super-resolution imaging of Taurus-Auriga protoplanetary disks: Probing statistical properties of disk substructures}",
      journal = {\pasj},
         year = 2024,
        month = jun,
       volume = {76},
       number = {3},
        pages = {437-474},
          doi = {10.1093/pasj/psae022},
archivePrefix = {arXiv},
       eprint = {2404.13570},
 primaryClass = {astro-ph.EP},
       adsurl = {https://ui.adsabs.harvard.edu/abs/2024PASJ...76..437Y}
}
\bibliographystyle{aasjournal}



\end{document}